\documentclass[%
 reprint,
 amsmath,amssymb,
 aps,superscriptaddress
]{revtex4-2}

\usepackage{graphicx}
\usepackage{dcolumn}
\usepackage{bm}
\usepackage{graphicx}
\usepackage{epstopdf}
\usepackage{float}
\usepackage{color}
\usepackage{cases}

\begin{document}
 \vfill

\title{Trapping polar molecules by surface acoustic waves}

\author{Haijin Ding}
\affiliation{School of Integrated Circuits, Tsinghua University, Beijing 100084, China}
\affiliation{Laboratoire des Signaux et Syst\`{e}mes (L2S), CNRS-CentraleSup\'{e}lec-Universit\'{e} Paris-Sud, Universit\'{e} Paris-Saclay, 3, Rue Joliot Curie, 91190, Gif-sur-Yvette, France}
\author{Re-Bing Wu}
\affiliation{Center for Intelligent and Networked Systems (CFINS), Department of Automation, Tsinghua University, Beijing 100084, China}
\author{Yu-xi Liu}\email{yuxiliu@mail.tsinghua.edu.cn}
\affiliation{School of Integrated Circuits, Tsinghua University, Beijing 100084, China}

\begin{abstract}
We propose a method to trap polar molecules with the electrical force induced by the surface acoustic wave (SAW) on piezoelectric materials. In this approach, the electrical force is perpendicular to the moving direction of the polar molecules, and is used to control the positions of trapped polar molecules in the direction orthogonal to the acoustic transmission. By virtue of an external electrical force, the SAW-induced electrical field can trap the polar molecules into single-layer or multi-layer lattices. The arrangement of molecules can affect the binding energy and localization of the molecule array. Then the one- or two-dimensional trapped polar molecule arrays can be used to construct the Bose-Hubbard (BH) model, whose energy and dynamics are affected by the localizations of the trapped molecules. We find that the phase transitions between the superfluid and Mott insulator based on trapped polar molecule BH model can be modulated by the SAW induced electrical potential.
\end{abstract}


\maketitle

\section{\label{sec:level1}Introduction}
{Polar} molecules have many potential applications such as in quantum computation~\cite{carr2009cold,PMquantumComputation,demille2002quantum,PMqudit,PMNaturePhysics,yelin2006schemes} and simulations~\cite{RMPPolarM,PMHubSimu,QuantumSimu,QchemistryReview}. The first step towards manipulating polar molecules is to decelerate and trap them for further operations~\cite{doyle2004quo}. It has been shown that polar molecules can be trapped by magnetic~\cite{MagneticTrapNature,YejunMagnet,MtrapDemille}, optical~\cite{caldwell2021general,zhang2020forming,liu2019molecular,liu2018building,friedrich2022electro} or electrical fields~\cite{meek2009trapping,YinJPThreeD,Meijerdeceleration,ElectrostaticTrapVelocity,TwofieldTrap,Actrapvan2005,friedrich2022electro}. For examples, in the magneto electrostatic
trap, polar molecules can be stably trapped by the magnetic quadrupole after deceleration~\cite{YejunMagnet}. By additionally applying both dc and ac electrical fields, the molecules can also be trapped via mechanical equilibrium~\cite{TwofieldTrap,bethlem2002deceleration,tarbutt2009prospects}. Once successfully trapped, the internal rovibronic state, center of mass and interactions among different polar molecules can further be modulated via additional control fields~\cite{charron2007quantum,rom2004state,micheli2006toolbox}. Variational efficient trapping approaches allow for more possible applications by controlling the trapped molecules~\cite{RMPPolarM,mitra2022quantum}.

The interaction between electrical fields and polar molecules can be studied via the Stark effect that the electrical field can modulate the spectral properties of polar molecules~\cite{hermansson1996electric,hochstrasser1973electric,hudson2006experiments,hudson2004efficient}. Based on this, polar molecules can be controlled or rotated by the electric field because of asymmetric structure and larger dipole moments, and this is different from the circumstance of neutral atoms~\cite{PMNaturePhysics}. The doublet splitting of polar molecules induced by the electric field can affect the Stark potential, which can further determine the mechanical potential, mechanical force, and mechanical motions of polar molecules~\cite{meekphdthesis,simpson2019control,DecelerationPRA,StarkHudson}.

Surface acoustic wave (SAW) devices have been extensively applied to the classical information processing~\cite{morgan2010surface,hashimoto2000surface}. The SAW, which is a kind of mechanical wave, within the piezoelectric materials is excited by an external ac voltage source upon the interdigital transducers (IDTs), which can transform the electrical energy to the mechanical energy in the form of propagating acoustic wave on the surface of the piezoelectric substrate or reversely transform the acoustic wave to the electrical energy.
The induced electrical potential by SAW can drive electrons to generate zero-resistance states~\cite{SAWelectrongas1}, acoustoelectrical currents~\cite{SAWelectrongas2}, and metal-Mott insulator phase transitions~\cite{SAWPhaseVelocity}.
With the tunable external ac source, the SAW in the piezoelectric materials can provide a well-controlled time-dependent moving electrical potential, and be designed to control electronic~\cite{Simon1996Coupling} or polar particles~\cite{frozen,SAWelectrongas2}.

In this paper, we propose a method to trap polar molecules using the electrical field in the free space carried by the SAW propagating along the surface of the piezoelectric substrate materials. To trap and control the polar molecule by the electrical force via its dipole moment~\cite{Yejun2017new,ni2010dipolarYeNature,PMNaturePhysics},
we consider that the surface acoustic wave can induce electrical potential both in the piezoelectric material and in the free space over electrodes~\cite{chen2007optimization,spectrum,Simon1996Coupling,Solution011}. This is similar to the circumstance that polar molecules can be modulated by the electrical forces applied by the electrodes with periodic voltages~\cite{meekPRL,meek2009trapping}.

The remainder of this paper is organized as follows. In Sec.~\ref{Sec:EnergyE}, we explore the possibility that polar molecules are trapped as one-, two- and three-dimensional arrays by the electrical field induced by IDTs of SAW  combined with another externally applied electrical force. 
In Sec.~\ref{Sec:ThreeDMultiLayer}, we study how the trapping approach can affect the property of the multi-layer trapped polar molecules. 
In Sec.~\ref{Sec:SpatialDynamics}, we clarify the spatial distribution of the single-layer trapped molecules. In Sec.~\ref{Sec:BHM}, we study the dynamics and phase transition of the Bose-Hubbard model based on polar molecules, which are trapped in lattices by SAW. Finally, conclusions are presented in Sec.~\ref{Sec:Conclusion}. 

\section{Theoretical model for trapping polar molecules}\label{Sec:EnergyE}
\begin{figure}[h]
\centerline{\includegraphics[width=1\columnwidth]{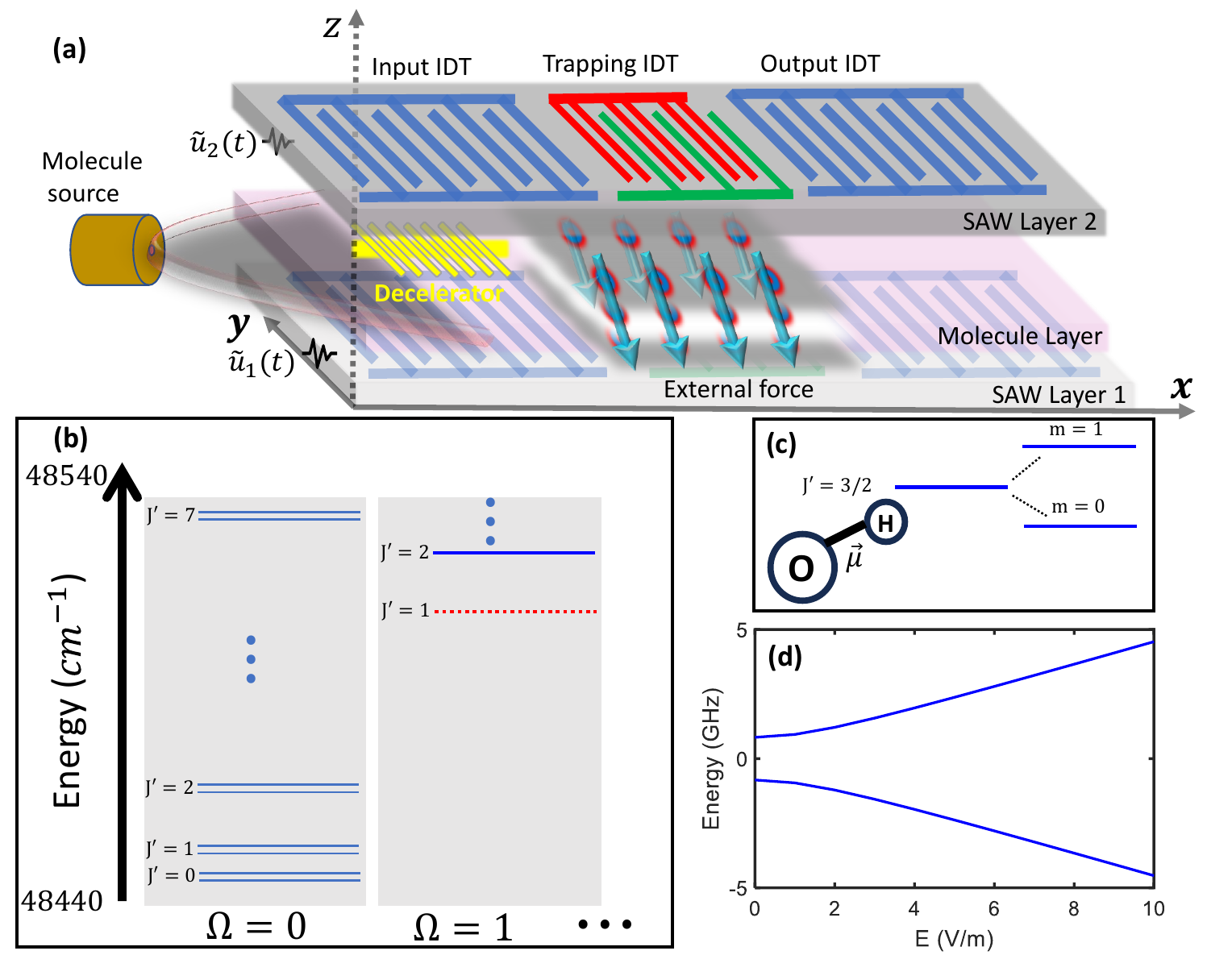}}
\caption{(a)~Schematic diagram for a setup of trapping polar molecules with SAW induced fields and the external electrical force. (b)~The energy level structure of CO with different quantum numbers~\cite{meek2009trapping,meekPRL}. (c)~The energy level structure of OH molecule with a stronger polarity~\cite{hudson2006experiments}. (d)~The energy level structure of OH molecule affected by the electrical field when $E_{\Lambda}=1.65$GHz, $|\vec{\mu}| = 1.67$Debye valued by Debye length, $J'=3/2$ and $m = \Omega = 1$~\cite{hudson2006experiments}.}	\label{fig:TraptopDesign} 
\end{figure}

As schematically shown in Fig.~\ref{fig:TraptopDesign}(a), polar molecules transmit between two layers of piezoelectric materials after being slowed down by the decelerator. In each piezoelectric layer, both the input and output IDTs consist of two comb-shaped arrays of metallic (i.e., the blue electrodes in two SAW layers in Fig.~\ref{fig:TraptopDesign}(a)). The primary functions of IDTs are to convert electric signals to SAW, or convert SAW back to electrical signal via piezoelectric effect. Driven by the external electrical fields $\tilde{u}_1(t)$ and $\tilde{u}_2(t)$, the input IDT excites acoustic waves and the output IDT converts SAW into electrical signals. In the middle trapping IDT, the three electrodes (i.e., two red electrodes and one green electrode in the upper layer of Fig.~\ref{fig:TraptopDesign}(a)) in each unit can transfer the mechanical oscillation of the SAW to the electrical field in the open space, during which the propagation of the acoustic wave remains uninfluenced by canceling out the reflection waves in the piezoelectric material~\cite{AlekseyPRL,morgan2010surface}. Besides electrical field from the trapping IDT, external designed electrical fields, represented with the blue down arrows in Fig.~\ref{fig:TraptopDesign}(a), can also be applied to the space between two layers of SAWs. The external electrical fields can be realized and modulated by a three-dimensional electrode array as in Ref.~\cite{Yin2022driven} or Appendix~\ref{sec:threeDelelctric}. Additionally, to enhance the efficiency of interactions, the two piezoelectric layers face to each other with the electrodes exposed to the open space between them. Then the energy of polar molecules, i.e., CO with the energy structure as in Fig.~\ref{fig:TraptopDesign}(b) or hydroxyl radical (OH) as in Fig.~\ref{fig:TraptopDesign}(c) can be affected by the electrical field by SAW as follows.
\subsection{Interaction between polar molecules and the electrical field by SAW}
According to the calculations shown in Appendix~\ref{sec:calWave1}, in the free space between two piezoelectric layers, the electrical field converted by the trapping IDT in the middle part of each layer can be represented as
\begin{small}
\begin{equation} \label{con:ExyvectorSimpleMain}
   \begin{cases}
   E_{x}^{(j)}(x,z,t)&=M\bar{u}_j k e^{-k\tilde{z}_j}  \sin\left[k\left(x-\frac{2\lambda}{3}-vt\right)\right] ,\\
   E_z^{(j)}(x,z,t)&=(-1)^{j-1} M \bar{u}_j k e^{-k\tilde{z}_j} \cos \left[k\left(x-\frac{2\lambda}{3}-vt\right)\right],
   \end{cases}
\end{equation}
\end{small}%
where $j=1,2$ represent the electrical fields induced by the lower and upper SAW layer respectively, $x$ represents the propagation direction of the acoustic wave, $z$ represents the vertical direction to the surface distinguished by $(-1)^{j-1}$ for the positive and negative directions, $M$ is the number of units for the trapping IDT (i.e., $M=3$ in Fig.~\ref{fig:TraptopDesign}(a), $\bar{u}_j$ is the amplitude of the voltage between the red and green electrodes of the trapping IDT, $\lambda$ is the wavelength of the acoustic wave that is equal to the periodicity of IDT stripes~\cite{AlekseyPRL}, $\tilde{z}_j$ is the distance to the lower or upper SAW layer with $\tilde{z}_1 = z$ and $\tilde{z}_2 = D-z$ where $D$ is the distance between two layers of piezoelectric materials, $v$ is the velocity of the acoustic wave and $k$ is the wave number. More details on the architecture design, the realization of external electrical force represented by the down arrows in Fig.~\ref{fig:TraptopDesign}(a) and the solutions of the acoustic waves are given in Appendix~\ref{sec:calWave1}.

When there is not external electrical force applied to the polar molecule, the molecule dynamics is governed by the Hamiltonian $H_0 = B\mathbf{J}^2$, where $\mathbf{J}$ is the rotational angular momentum and $B$ is the rotational constant~\cite{chen2010qubit}. When the polar molecules are over the IDT array of the surface acoustic wave, 
the electric field induced by the SAW in the $j$th layer can be coupled to the polar molecule via electric-dipole interactions, described by the Hamiltonian as~\cite{RMPPolarM,chen2010qubit}
\begin{small}
\begin{equation} \label{con:HiPMField}
\begin{aligned}
H_I= -\vec{\mu}\cdot\vec{E}^{(j)}(x,z,t),
\end{aligned}
\end{equation}
\end{small}%
where $\vec{\mu}$ is the dipole moment of the polar molecule, and the field $\vec{E}^{(j)}(x,z,t) = \left[E_{x}^{(j)}(x,z,t),E_{z}^{(j)}(x,z,t)\right]^T$ given in Eq.~(\ref{con:ExyvectorSimpleMain}) is from the trapping IDTs (see for more details in Appendix~\ref{sec:calWave1}). When the electric field by SAW is weak, the polar molecules (i.e., OH in Ref.~\cite{hudson2006experiments}, KCs in Ref.~\cite{demille2002quantum}, CaBr in Ref.~\cite{PMNaturePhysics}, deuterated
ammonia ($\mathrm{ND_3}$) in Ref.~\cite{ElectrostaticTrapVelocity}, CO in Refs.~\cite{meek2009trapping,meekPRL,yelin2006schemes}) can be regarded as a two-state system. The electric field along different directions has different coupling strengths with the dipole moments even for a given quantum number~\cite{meekphdthesis}. Here we consider the case that the dipole moments of polar molecules are along the $z$-direction, then Eq.~(\ref{con:HiPMField}) is further written as~\cite{meek2009trapping,meekPRL,meekNJP,meekphdthesis}
\begin{footnotesize}
\begin{equation} \label{con:Hsplit}
\begin{aligned}
H =  \begin{bmatrix}
    \omega_2 & -|\vec{\mu}| E_z^{(j)}(x,z,t)\frac{m\Omega}{J'\left(J'+1\right)} \\
     -|\vec{\mu}| E_z^{(j)}(x,z,t)\frac{m\Omega}{J'\left(J'+1\right)} & \omega_1\\
  \end{bmatrix},
\end{aligned}
\end{equation}
\end{footnotesize}%
where we consider two energy levels of the polar molecule as $\hbar \omega_l$ with $l=1,2$, $\omega_2 - \omega_1 = \bar{E}_{\Lambda}$ is the $\Lambda$-doublet splitting, $E_z^{(j)}(x,z,t)$ is given in Eq.~(\ref{con:ExyvectorSimpleMain}), $J'$, $\Omega$ and $m$ are the quantum numbers according to the energy level structure of a polar molecule, i.e.,  $J'$ is the index of the energy level, $\Omega$ and $m$ with the values $-1,0,1$ represent quantum numbers for the basis state of the molecule wave function~\cite{meekphdthesis}.

In the rotating reference frame (RRF) with the rotating wave approximation (RWA)~\cite{hughes2020robust,tai1988ac}, the Hamiltonian representing the interaction between the polar molecule and electric field can be simplified as 
\begin{small}
\begin{equation} \label{con:Hsimple}
\begin{aligned}
H' =\begin{bmatrix}
    \bar{\omega} + \frac{\bar{E}_{\Lambda}-kv}{2}  &(-1)^{j} \frac{|\vec{\mu}|m\Omega }{2J'\left(J'+1\right)}\mathcal{E}_x^*\bar{E}_z^{(j)} \\
     (-1)^{j} \frac{|\vec{\mu}|m\Omega }{2J'\left(J'+1\right)} \mathcal{E}_x \bar{E}_z^{(j)}  & \bar{\omega} - \frac{\bar{E}_{\Lambda}-kv}{2}\\
  \end{bmatrix},
\end{aligned}
\end{equation}
\end{small}%
where $\bar{E}_z^{(j)} = M \bar{u}_j k e^{-k\tilde{z}_j}$, $\mathcal{E}_x = e^{-ik\left(x-\frac{2\lambda}{3}\right)} $, $\bar{\omega} = \left(\omega_1 + \omega_2 \right)/2$, and more details are given in Appendix~\ref{sec:ApendixRWA}.
Then the energy levels of the polar molecule affected by the electric field produced by the SAW in the $j$th layer are 
\begin{equation} \label{con:Uz}
\begin{aligned}
\mathbf{E}_j&= \bar{\omega}  \pm \sqrt{\left(\frac{E_{\Lambda}}{2}\right)^2 + \mu_{\rm eff}^2 M^2\bar{u}_j^2 k^2 e^{-2k\tilde{z}_j}}\\
&\equiv \bar{\omega}  + U_j,
\end{aligned}
\end{equation}%
where $E_{\Lambda} = \bar{E}_{\Lambda} - kv$, 
$U_j = \pm \sqrt{\left(E_{\Lambda}/2\right)^2 + \mu_{\rm eff}^2 \left| \bar{E}_z^{(j)}  \right|^2}$
representing the modulation of energy levels by the SAW provided electrical field, the effective dipole strength $\mu_{\rm eff} = \pm|\vec{\mu}| m\Omega/2J'\left(J'+1\right)$, which can be positive for the high-field seeking states or negative for the low-field seeking states~\cite{wohlfart2008stark,meekphdthesis}. Obviously, $\mathbf{E}_j$ is independent of $\mathcal{E}_x$ in the RRF. In the following, we take $M=1$ for simplification.

\begin{figure}[h]
\centerline{\includegraphics[width=1\columnwidth]{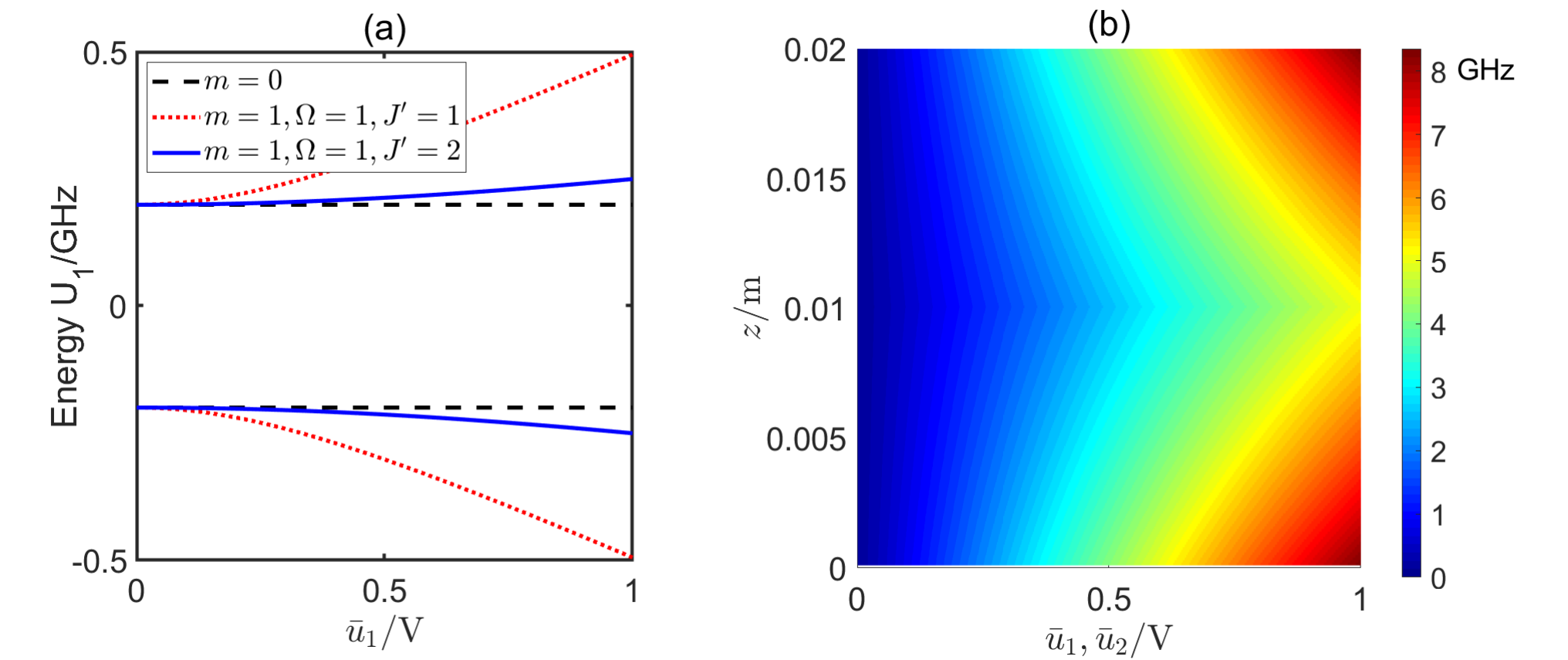}}
\caption{(a)~The dependence of the energy levels of polar molecules on the electrical field with different quantum numbers, where the distance between the molecules and the lower SAW layer is $z = 0.01$m, $E_{\Lambda} = 0.4$GHz, $|\vec{\mu}| = 0.167$Debye and the wave number $k =10$. (b)~The energy level of polar molecules CO evaluated by $\max\left(U_j\left (\bar{u}_j,z\right )\right)$ with $j=1,2$ representing the lower or upper SAW when $k=50$, $E_{\Lambda} = 0.4$GHz, $|\vec{\mu}| = 0.167$Debye, $J' =m=\Omega =1$~\cite{meekphdthesis}.}
	\label{fig:EnergyLevel}
\end{figure}

The energy levels of the polar molecule in Fig.~\ref{fig:TraptopDesign}(d) are with the parameters in Ref.~\cite{hudson2006experiments} and in Fig.~\ref{fig:EnergyLevel} with the parameters in Refs.~\cite{meek2009trapping,meekphdthesis,meekPRL}, which are affected by the electrical field along the $z$ direction. We find that the energy gap strongly depends on the quantum number and the amplitude of voltage.

\subsection{Trapping polar molecules into one-, two- and three-dimensional lattices}
If we want to trap the polar molecule at $\vec{r} = [x_0,z_0]$, the joint electrical force applied to the polar molecule $\vec{F}(x_0,z_0,t)$ must satisfy the following conditions~\cite{Trapcondition,PMactrap,Actrapvan2005}
\begin{small}
\begin{equation} \label{con:forcecondition}
\begin{cases}
\vec{F}(x_0,z_0,t) \equiv 0,\\
\vec{\nabla} \cdot \vec{F}(x_0,z_0,t) < 0,
\end{cases}
\end{equation}
\end{small}%
where the first line is for the mechanical equilibrium, and the second line represents the occurrence of resilience effect once the molecules escape from the trapped site $\left ( x_0,z_0\right)$.

As a component of $\vec{F}(x_0,z_0,t)$, the electrical force from the trapping IDT of the $j$th SAW layer can be represented as $\vec{F}^{(j)}(x,z,t) = \left [F_x^{(j)}(x,z,t), F_z^{(j)}(x,z,t)\right]^T$ when the molecule moves slowly along the $x$-direction after deceleration, 
and $\vec{F}^{(j)}(x,z,t)$ is determined by the strength of the electrical field from the trapping IDTs and the quantum number of the polar molecule. With the energy $U_j$ in Eq.~(\ref{con:Uz}), the electrical force produced by SAW upon polar molecules reads~\cite{meek2009trapping}
\begin{small}
\begin{equation} \label{con:forceCO}
\begin{aligned}
\vec{F}^{(j)}(x,z,t) &= -\frac{d U_j}{d\left|\bar{E}_z^{(j)}\right|} \vec{\nabla} \left|\bar{E}_z^{(j)}\right| \\
&= -\frac{1}{2}\frac{1}{\left |\bar{E}_z^{(j)}\right |} \frac{dU_j}{d\left|\bar{E}_z^{(j)}\right|} \vec{\nabla} \left|\bar{E}_z^{(j)}\right|^2,\\
\end{aligned}
\end{equation}
\end{small}%
where

$d U_j/d\left|\bar{E}^{(j)}_z\right| = \mu_{\rm eff}^2\left|\bar{E}_z^{(j)}\right|/\sqrt{\left(E_{\Lambda}/2\right)^2 + \left(\mu_{\rm eff}\left|\bar{E}_z^{(j)}\right|\right)^2}$,
then
\begin{small}
\begin{subequations} \label{con:forceCOxz}
\begin{numcases}{}
F_x^{(j)}(x,z,t) = 0,\label{Fxdirect}\\
F_z^{(j)}(x,z,t) =\frac{\mu_{\rm eff}^2  \bar{u}_j^2k^3e^{-2k\tilde{z}_j}}{\sqrt{\left(\frac{E_{\Lambda}}{2}\right )^2 + \left (\mu_{\rm eff}\left|\bar{E}_z^{(j)}\right|\right )^2}}.\label{Fzdirect}
\end{numcases}
\end{subequations}
\end{small}%

Additionally, we can also have
\begin{footnotesize}
\begin{equation} \label{con:condition2}
\begin{aligned}
\vec{\nabla} \cdot \vec{F}^{(j)}(x,z,t)
=- \frac{\mu_{\rm eff}^2\left[\left (\frac{\partial^2 \phi}{\partial x^2} \right )^2 + \left (\frac{\partial^2 \phi}{\partial z^2} \right )^2 + 2 \left ( \frac{\partial^2 \phi}{\partial z\partial x} \right )^2\right]}{\sqrt{\left (\frac{E_{\Lambda}}{2}\right )^2 + \left (\mu_{\rm eff}\left|\vec{E}^{(j)}\right|\right )^2}}<0,
\end{aligned}
\end{equation}
\end{footnotesize}%
where $\phi = \phi_j(x,z)$ is the electrical potential produced by trapping IDTs in the $j$th SAW layer and more details are given in Appendix~\ref{sec:ProofEq}. Equation~(\ref{con:condition2}) indicates that the electrical field induced by SAW can trap the polar molecules if the condition of the equilibrium in Eq.(\ref{con:forcecondition}) can be satisfied. 

\begin{figure}
\centerline{\includegraphics[width=1\columnwidth]{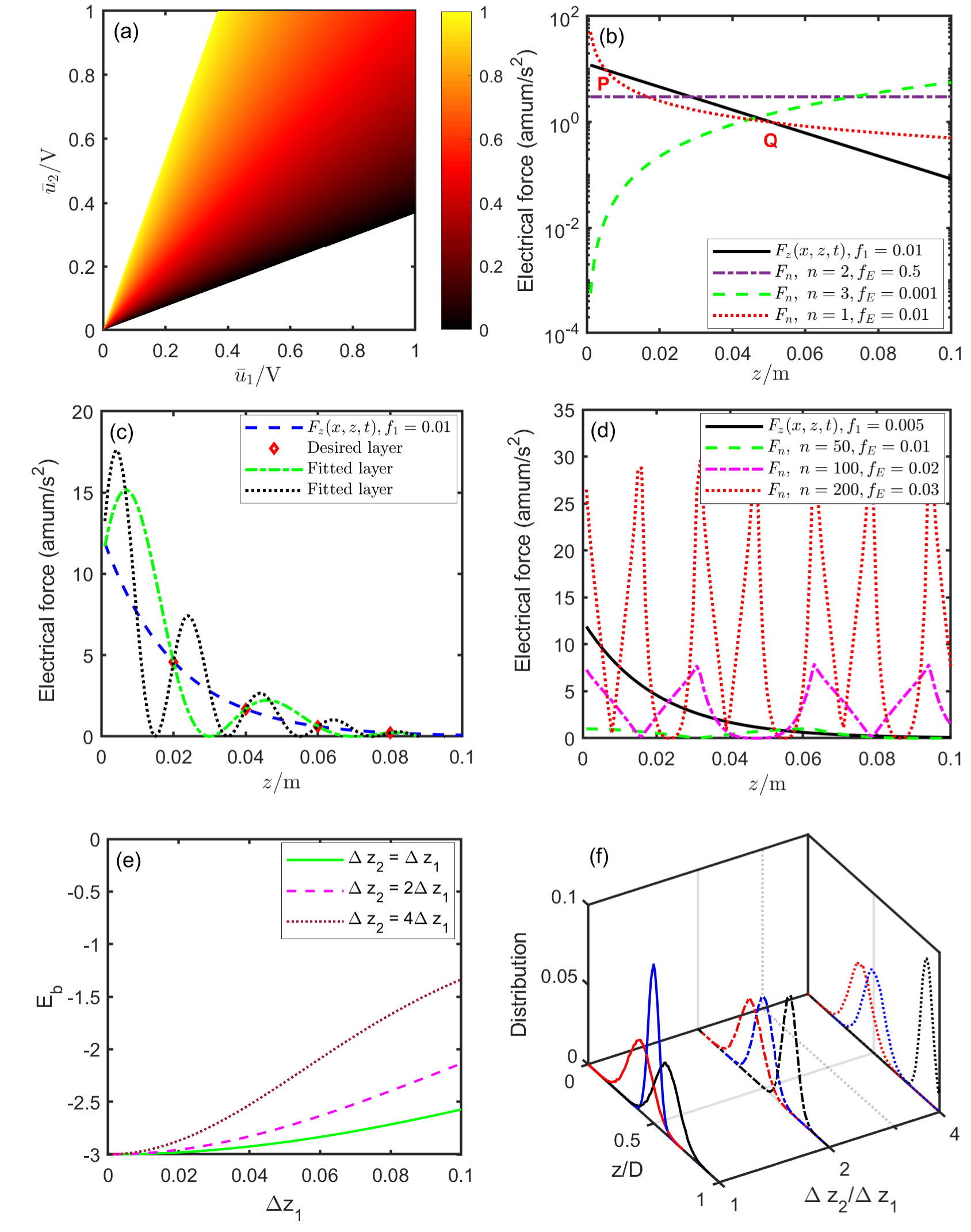}}
\caption{(a)~Trapping locations with two layers of surface acoustic waves, where $k =50$, $D = 0.02m$ and $z_0/D$ are evaluated. The blank area represents that the molecules cannot be trapped for values of $\bar{u}_1$ and $\bar{u}_2$. (b)~$F_z(x,z,t)$ represents the force of the lower SAW with different locations of $z$, the dash or dot curves represent the force of different external electrical field represented as $\tilde{E}(z) = f_E z^n$. The electrical forces are valued by the unit amu$m/s^2$ as in Ref.~\cite{Yin2022driven} with amu representing the atomic mass unit~\cite{wangLukinPRL}. When the dash curve and the solid curve intersect once, it means that the molecules can be only trapped in the single layer with the value of $z$ at the crossover point. When the dot curve and the solid curve intersect twice as the red curve, the polar molecules can be trapped in two layers at the two crossover points.
(c)~Green-dash represents the electric field $\tilde{E}(z)= (- 2.42\times10^4z^3 +5201z^2- 389.6z+ 10.5)\left ( 1+\sin\left (\pi z/0.02\right ) \right )$ with an envelope of a sine function to trap within the desired layers, and the trapping by the black-dot curve with a cosine envelop is more stable because the joint force has different restoring directions below or over the desired layers. 
(d)~Nonuniform trapping of polar molecules with the external electrical field $\tilde{E}(z) = f_E \sin(nz)$ with different values of $n$ and $f_E$, and the force upon the polar molecule reads $F_n =\left |nf_E^2\sin(nz)\cos(nz)/\left (|f_E \sin{nz}|+\tilde{\epsilon}\right ) \right |$ where we take $\tilde{\epsilon} = \sqrt{\left(E_{\Lambda}/2/\mu_{\rm eff}\right)^2 + |\tilde{E}(z)|^2} -  |\tilde{E}(z)| \approx 0.01$ and is close to zero~\cite{meekphdthesis}.
(e)~Comparison of the binding energy with different trapped locations of layers with $\alpha= 8$. (f)~The distribution of trapped locations of polar molecules along the $z$ direction in different layers, which can be valuated with $R_j$ in Eq.~(\ref{con:Rj}) with $R_0 = 0.04$, $\xi = 0.01m^2$, $z_2 = 0.5D$ and $\Delta z_2 = 0.4D$. The parameters in (e) and (f) are chosen according to Refs.~\cite{wangLukinPRL,weaktrap}.
}
	\label{fig:Trapresult}
\end{figure}

When $\bar{u}_1$ and $\bar{u}_2$ are nonzero and there are no external forces, the polar molecules can be trapped by the electrical forces of IDTs between two piezoelectric materials, and the trapped location $z_0$ of molecules is determined by the equilibrium with $\bar{u}_1^2 e^{-2kz_0} = \bar{u}_2^2 e^{-2k(D-z_0)}$, where $\bar{u}_1$ and $\bar{u}_2$ are proportional to the amplitudes of $u_1(t)$ and $u_2(t)$ in Fig.~\ref{fig:TraptopDesign}(a), respectively. Then we can obtain the trapping location $z_0 = \ln \left (e^{kD}\bar{u}_1/\bar{u}_2 \right )/2k$, the relationship $e^{-kD}<\bar{u}_1/\bar{u}_2<e^{kD}$ always holds because $0<z_0<D$, and this agrees with the fact that larger $D$ induces more robust parameter settings of $\bar{u}_1$ and $\bar{u}_2$ for trapping the molecules. As shown in Fig.~\ref{fig:Trapresult}(a), the trapping locations are closer to the lower SAW layer when $\bar{u}_1>\bar{u}_2$ and are closer to the upper SAW layer when $\bar{u}_2 > \bar{u}_1$. In this way, a single layer of polar molecules can be trapped at arbitrary locations of the height $z_0$. Besides, here the potential applied on the polar molecules is the joint potential generated by the trapping IDTs of the upper and lower SAW layers. Then Eq.~(\ref{con:condition2Derive}) in Appendix~\ref{sec:ProofEq} reveals that the molecules can be stably trapped at the equilibrium positions.

When $\bar{u}_1 \neq 0$ and $\bar{u}_2 = 0$, the polar molecules can be trapped at $(x_0,z_0)$ by the electrical field induced by the trapping IDT in the lower SAW layer and the external electrical force $\tilde{F}_z(x_0,z_0,t)$, which satisfies the equilibrium condition, that is 
\begin{footnotesize}
\begin{equation} \label{con:forceCosine}
\begin{aligned}
\tilde{F}_z(x_0,z_0,t) &=F_z^{(1)}(x_0,z_0,t)\\
 &= \frac{\mu_{\rm eff}^2 M^2\bar{u}_1^2k^3e^{-2kz_0}}{\sqrt{\left (E_{\Lambda}/2\right )^2 + \left (\mu_{\rm eff}\left|\bar{E}_z^{(1)}\right|\right )^2}} \\
&\triangleq f_1 k^2 e^{-kz_0},
\end{aligned}
\end{equation}
\end{footnotesize}%
with $f_1 \approx \mu_{\rm eff} M \bar{u}_1 $. $\tilde{F}_z(x_0,z_0,t)$ can be used to cancel $F_z^{(1)}(x_0,z_0,t)$ based on the approximation $E_{\Lambda} \ll  \left|\bar{E}^{(1)}(x,z,t)\right|$~\cite{frozen}.

The external electrical force $\tilde{F}_z(x_0,z_0,t)$ can be realized with the external electric field $\tilde{E}(z)$ applied upon the polar molecule array, then the mechanical equilibrium along $z$ direction can be realized when~\cite{meekphdthesis} 
\begin{small}
\begin{equation} 
\begin{aligned}
\frac{1}{2|\tilde{E}(z)|}\frac{\partial |\tilde{E}(z)|^2}{\partial z}= f_1 k^2 e^{-kz},
\end{aligned}
\end{equation}
\end{small}%
by representing $\tilde{F}_z(x_0,z_0,t)$ in Eq.~(\ref{con:forceCosine}) with the format of $\tilde{E}(z)$, and the equation is independent of $x$.
Mathematically, different $\tilde{E}(z)$ can induce different trapping results for polar molecules. For examples, when $D = 10$cm and $k=50$, various trapping approaches are compared in Figs.~\ref{fig:Trapresult} (b-d).

Different from trapping with two SAW layers where the molecules can be trapped at arbitrary horizontal locations at $z_0$ with unlimited number of trapped molecules, the polar molecules can only be trapped at the location where external force for the equilibrium exists as in Fig.~\ref{fig:TraptopDesign}(a) when only one SAW layer is used for trapping.
Then the molecules can be trapped in a lattice with external electrical force at the height $z$ determined by the one or several different intersections as in Figs.~\ref{fig:Trapresult}(b-d).

Further for the simulations in Fig.~\ref{fig:Trapresult}(b), the red curve denoting external electric force has two intersection points $\mathrm{P}$ and $\mathrm{Q}$ with the SAW induced electrical force. For the lower intersection point $\mathrm{P}$ at the height $z_p$, the combined force is downward when $z<z_p$ and upward when $z>z_p$, thus the trapped polar molecules can easily escape from the trapped site at $z_p$. However, for the upper intersection point $\mathrm{Q}$ at the height $z_q$, the combined force is upward when $z<z_q$ and downward when $z>z_q$, and this provides a resilience along $z$ direction to make sure that the polar molecules are stably trapped at $z_q$. This is why in Fig.\ref{fig:Trapresult}(c), the black-dot curve can trap the polar molecules at the desired layers more stably than the green-dash curve. Although the black-dot curve can simultaneously introduce unstable trapped intersections, while the number of trapped polar molecules is much less than that by the stable intersections.

Once the polar molecules are stably trapped by the electrical force, the interactions among molecules are affected by their locations and external electrical potentials. In the following, we discuss three different cases: the first is that the polar molecules are trapped as a three-dimensional array, the lattice sites in each layer are remotely separated in the horizontal direction and the horizontal hoppings are not considered, the second is that a single layer of polar molecules are trapped and the horizontal hoppings are considered, and the third is that the polar molecules are trapped by one SAW layer (i.e., the upper layer) and the external force in lattices, but their quantum dynamics are also affected by the electrical potential induced by the other SAW layer, so that the lattice dynamics can be influenced by the lower SAW.

\section{Three-dimensional trapped polar molecule array with multiple layers}  \label{Sec:ThreeDMultiLayer}
When the polar molecules with three-dimensional structure are trapped in different heights as schematically shown in Fig.~\ref{fig:TraptopDesign}(a), the number of trapped molecules in each layer is the same because the external electrical forces $F_n$ are applied along the $z$ direction coming across each molecule layer, where $F_n$ is similarly calculated as $\vec{F}^{(j)}(x,z,t)$ in Eq.~(\ref{con:forceCO}) by replacing $\bar{E}_z^{(j)}$ with $\tilde{E}(z)$ under the approximation $E_{\Lambda} \ll \max \left|\tilde{E}(z)\right|$. For example, in Fig.~\ref{fig:Trapresult}(b) when $n=3$ and $f_E = 0.001$, the polar molecules can be trapped in a single layer that corresponds to the intersection point between the black and green curves.
By multiplying $\tilde{E}(z)$ with an envelope function, i.e., $f_{ev}= 1+\prod_{j=1}^N(z-z_j)$ or $f_{ev} = 1+\sin\left (\pi z/\Delta_z \right )$ with $\Delta_z = 0.02$ as shown in Fig.~\ref{fig:Trapresult}(c), the polar molecules can be evenly (green dash) or not evenly (Fig.~\ref{fig:Trapresult}(d)) trapped at the designed heights with $z_1,z_2,\cdots,z_N$. The method can also be generalized to the case of $\bar{u}_1 = 0$ and $\bar{u}_2\neq 0$ representing that molecules can be trapped with the upper SAW layer and the external electrical field.

When the polar molecules are trapped as a three-dimensional array, the attractive interactions among molecules in different trapped layers can bind them into chains along $z$ direction, and the molecular attraction can be evaluated with the binding energy dependent on the number and heights of trapped layers. The Hamiltonian of the trapped polar molecules along $z$ direction reads~\cite{wangLukinPRL}
\begin{small}
\begin{equation} \label{con:Hpm}
\begin{aligned}
H_{L} = \sum_{l=1}^L  \left( \frac{P_l^2}{2m_0} + \frac{m_0\omega^2r_l^2}{2} \right) + \frac{1}{2}\sum_{q\neq l}^L V_{|q-l|}\left(|z_i-z_l|\right),
\end{aligned}
\end{equation}
\end{small}%
where $m_0$ is the mass of the trapped polar molecules in each layer, $L$ is the number of trapped layers, $P_l$ and $r_l$ are the momentum and the position of the center-of-mass of the molecules in the $l$th layer, respectively, $\omega \propto k$ is the trapping frequency of the molecules, $V_{|q-l|}$ is the dipole-dipole interaction between molecules in the $q$th and $l$th layers, and is determined by the trapped height $z_q$ and $z_l$ respectively. The variational wave function of the $L$ layers of trapped polar molecules is~\cite{wangLukinPRL}
\begin{small}
\begin{equation} \label{con:Wavefunction}
\begin{aligned}
\psi_L(r_1,\cdots,r_L) = \prod_{l=1}^L\frac{\exp(-|r_l|^2/2R_l^2)}{\sqrt{\pi}R_l},
\end{aligned}
\end{equation}
\end{small}%
where the standard deviation $R_l$ of the normal distribution is determined by the equilibrium of the attractive forces between the polar molecules in the layers on its upper and lower sides. When the molecules are uniformly trapped along the $z$-direction as in Ref.~\cite{wangLukinPRL}, $R_l$ is the smallest at the middle layer. However, when the polar molecules are not uniformly trapped, $R_l$ should be proportional to $1+\xi|F_{\uparrow}^l - F_{\downarrow}^l|$ where $F_{\uparrow}^l = \sum_{q=l+1}^L \mathrm{d}V_{|q-l|}/\mathrm{d} |z_q-z_l|$ is the attractive force by the upper trapped polar molecules, $F_{\downarrow}^l = \sum_{q=1}^{l-1} \mathrm{d}V_{|q-l|}/\mathrm{d} |z_q-z_l|$ is the attractive force by the lower trapped polar molecules, and $\xi$ is a chosen parameter. Then, we have
\begin{small}
\begin{equation} \label{con:Rj}
\begin{aligned}
R_l 
=R_0\left(1+\xi \left|\sum_{q=l+1}^L \frac{1}{|z_q-z_l|^2} - \sum_{q=1}^{l-1} \frac{1}{|z_q-z_l|^2}\right|\right),
\end{aligned}
\end{equation}
\end{small}%
as a generalization of the formula in Ref.~\cite{wangLukinPRL}, and the parameter $R_l$ is determined by the trapped locations of different polar molecule layers. For the multi-layer trapped molecules, the binding energy representing the required energy to separate the layers reads $E_b = \sum_{q\neq l}E_b^{ql}\approx-\sum_{q\neq l}\exp(-\alpha(z_q-z_l)^2)$~\cite{weaktrap}.

Consider the case that three molecules layers are trapped at the heights $z_1<z_2<z_3$ as an example. Denote $\Delta z_1 = z_2 - z_1$ and $\Delta z_2 = z_3 - z_2$. The comparisons of the binding energy and the oscillations of polar molecules in different layers are shown in Figs.~\ref{fig:Trapresult}(e) and (f) with the parameters chosen according to Ref.~\cite{wangLukinPRL}. The absolute value of the binding energy for the multi-layer polar molecule array is larger when the molecules are closely trapped along the $z$ direction. When the layers are uniformly trapped with $\Delta z_1 = \Delta z_2$, the oscillation of the central layer is reduced. However, when the layers are nonuniformly trapped in dispersed locations, the oscillation of polar molecules is stronger due to the non-equalizing of the attractive force along the $z$ direction. Besides, the distributions of polar molecules in different layers are influenced by the thermal effect, as introduced in detail in Ref.~\cite{wangLukinPRL}.

\section{Spatial distribution of trapped molecules} \label{Sec:SpatialDynamics}
In a single trapped polar molecule layer, the polar molecules can hop among different localized lattice sites via quantum jumps, which can be called the Anderson localization~\cite{anderson1958absence,xu2015quantumAL}. When the molecules are trapped by the lower SAW as shown in Fig.~\ref{fig:TraptopDesign}(a) and an external electrical field as shown in Figs.~\ref{fig:Trapresult}(b-d), the molecules can be further trapped in 1D or 2D lattices, and the locations of the lattices are determined by the spatial distribution of the external electrical fields in the plane orthogonal to the $z$ axis. We denote the probability amplitude as $p_{\mathbf{n}}$ that a molecule is trapped at the lattice labeled with $\mathbf{n}$, where $\mathbf{n} = n$ when the molecules are trapped in the 1D lattice and $\mathbf{n} = \left (n_x,n_y\right )$ for the 2D lattice.
We use the one-dimensional trapped polar molecules along the $x$ axis as an example, the external electrical field for trapping is imposed at $x_1,x_2,\cdots,x_N$, the evolution of $p_n$ reads~\cite{anderson1958absence}
\begin{small}
\begin{equation} \label{con:pnevolution}
\begin{aligned}
\dot{p}_n(t) = -iU_0p_n(t) - i\sum_{j\neq n}v_{nj} p_j(t),
\end{aligned}
\end{equation}
\end{small}%
where $U_0$ is the energy of the polar molecules, $v_{nj}\propto \left | x_n - x_j\right|^{-3}$ represents the dipole interaction between the polar molecules trapped at the $n$th and $j$th lattices, and $1 \leq n,j \leq N$.

Applying the Laplace transform to Eq.~(\ref{con:pnevolution}), we have
\begin{small}
\begin{equation} \label{con:Pns0}
\begin{aligned}
P_n(s) &= \frac{p_n(0)- i\sum_{j\neq n}v_{nj} P_j(s)}{s+iU_0}\\
&=\frac{ip_n(0)}{is-U_0} + \sum_{j\neq n} \frac{1}{is-U_0}v_{nj} P_j(s),
\end{aligned}
\end{equation}
\end{small}%
where $P_j(s)$ is the Laplace transformation of $p_j(t)$.

Let us assume initially $p_1(0) = 1$ and $p_n(0) = 0$ for $n =2,3,\cdots,N$, that is, only molecules at the first lattice are initially trapped. Equation~(\ref{con:Pns0}) with $n=1$ reads~\cite{anderson1958absence}
\begin{small}
\begin{equation} \label{con:P1s}
\begin{aligned}
P_1(s) &= \frac{i}{is-U_0} + \sum_{j=2}^N \frac{1}{is-U_0}v_{1j} P_j(s),\\
\end{aligned}
\end{equation}
\end{small}%
and
\begin{small}
\begin{equation} \label{con:Pns}
\begin{aligned}
 P_n(s)&=\sum_{j\neq n} \frac{1}{is-U_0}v_{nj} P_j(s)\\
 &=\frac{1}{is-U_0}v_{n1} P_1(s) +  \sum_{j=2}^{n-1} \frac{1}{is-U_0}v_{nj} P_j(s)\\
 &~~~~+  \sum_{n+1}^{N} \frac{1}{is-U_0}v_{nj} P_j(s),
\end{aligned}
\end{equation}
\end{small}%
for $n=2,3,\cdots,N.$

Using Eq.~(\ref{con:Pns}), Eq.~(\ref{con:P1s}) becomes
\begin{small}
\begin{equation} \label{con:P1sCal2}
\begin{aligned}
&P_1(s) = \frac{i}{is-U_0} + \sum_{j=2}^N \frac{1}{is-U_0}v_{1j} \left [ \frac{1}{is-U_0}v_{j1} P_1(s) \right.\\
&\left. +  \sum_{j'=2}^{j-1} \frac{1}{is-U_0}v_{jj'} P_{j'}(s)+  \sum_{j+1}^{N} \frac{1}{is-U_0}v_{jj'} P_{j'}(s) \right ]\\
&=\frac{i}{is-U_0} + \frac{P_1(s)}{(is-U_0)^2}\sum_{j=2}^N v_{1j}^2 \\
& + \frac{1}{(is-U_0)^2}\sum_{j=2}^N v_{1j}\left [\sum_{j'=2}^{j-1} v_{jj'} P_{j'}(s)+  \sum_{j+1}^{N} v_{jj'} P_{j'}(s) \right ]\\
&\approx \frac{i}{is-U_0} + \frac{P_1(s)}{(is-U_0)^2}\sum_{j=2}^N v_{1j}^2 + \frac{1}{(is-U_0)^3}\sum_{j=2}^N v_{1j}\\
&\left [\sum_{j'=2}^{j-1} v_{jj'} v_{j'1} P_1(s) +  \sum_{j+1}^{N} v_{jj'}v_{j'1} P_1(s)  \right ],\\
\end{aligned}
\end{equation}
\end{small}%
where $P_{j'}(s) \approx \frac{1}{is-U_0}v_{j'1} P_1(s)$ has been used. Then
\begin{small}
\begin{equation} \label{con:P1sSolution}
\begin{aligned}
&P_1(s)\approx\frac{i}{is-U_0} +\frac{\sum_{j=2}^N v_{1j}^2}{(is-U_0)^2}P_1(s) \\
&+ \frac{1}{(is-U_0)^3} \sum_{j=2}^N v_{1j} \left (\sum_{j'=2}^{j-1} v_{jj'} v_{j'1}+  \sum_{j+1}^{N} v_{jj'}v_{j'1}\right )P_1(s).
\end{aligned}
\end{equation}
\end{small}%
Denoting $\alpha = \sum_{j=2}^N v_{1j}^2$ and $\beta = \sum_{j=2}^N v_{1j} \left (\sum_{j'=2}^{j-1} v_{jj'} v_{j'1}+  \sum_{j+1}^{N} v_{jj'}v_{j'1}\right )$, then we have
\begin{small}
\begin{equation} \label{con:P1salphaBeta}
\begin{aligned}
&P_1(s)= \frac{i(is-U_0)^2}{(is-U_0)^3-\alpha (is-U_0) - \beta}\\
& = \frac{i(is-U_0)^2}{-is^3 +3U_0s^2 +i(3U_0^2-\alpha)s - U_0^3 +\alpha U_0 -\beta}.
\end{aligned}
\end{equation}
\end{small}%
For most of the case $U_0^3 - \alpha U_0 + \beta \neq 0$, thus $\lim_{t\rightarrow \infty}p_1(t)= \lim_{s\rightarrow 0}sP_1(s) = 0$.

\begin{figure}[h]
\centerline{\includegraphics[width=1\columnwidth]{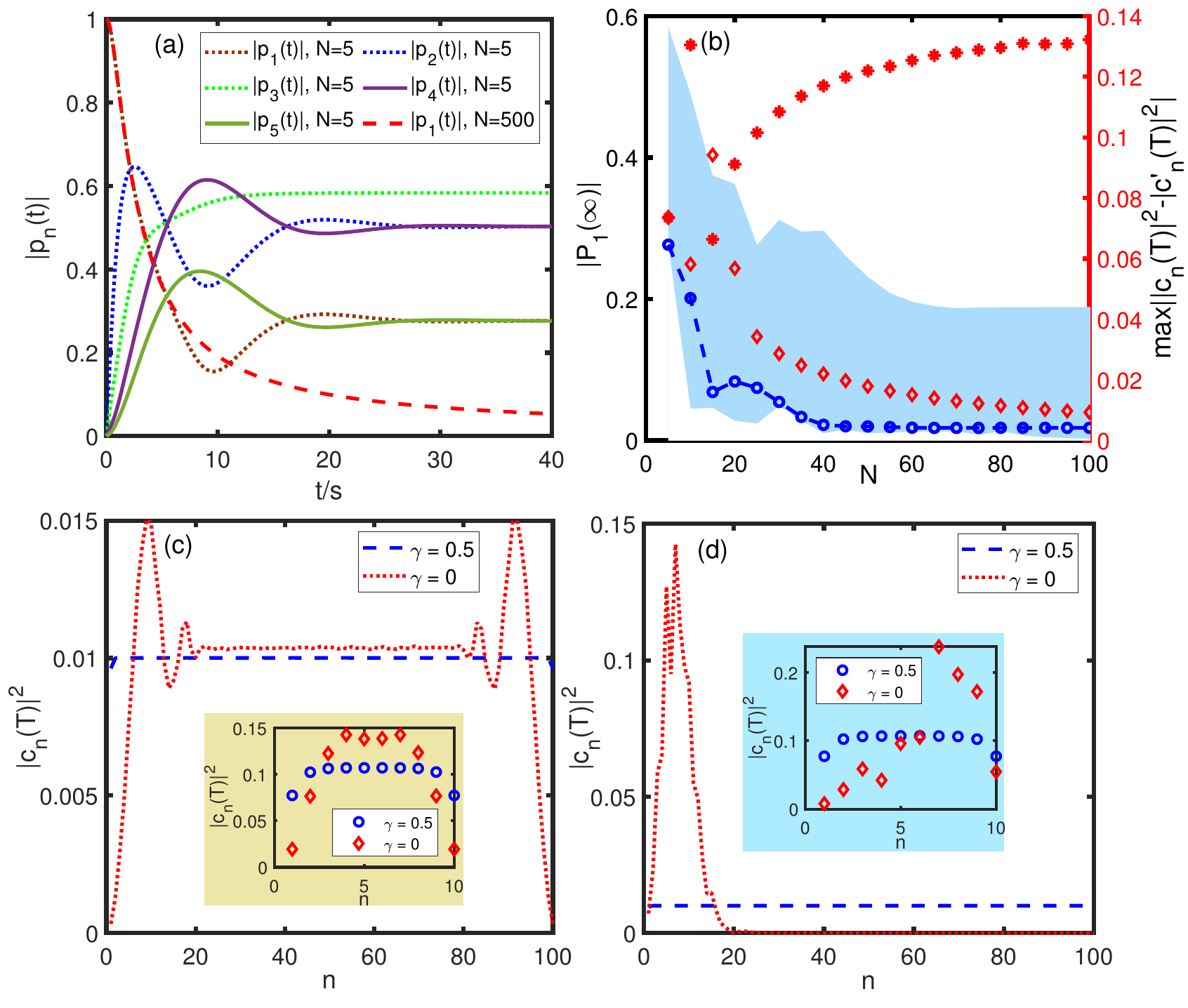}}
\caption{(a) Comparison of the probability of trapping at different localized sites with two different single-layer trapping methods. $U_0 = 2$ in both of simulations, $v_{nj} =  0.4\left | x_n - x_j\right|^{-3}$, $x_n = 0.1n$, $N=5$ for discrete trapping, and the discrete trapping converges to the continuous case when $N=500$. (b) The blue-dashed-circle evaluated by left-y axis represents the steady value of $P_1$ between $\max_n \left(P_n \right)$ and $\min_n \left(P_n \right)$ with the same parameters as (a), and decreases with the increase of lattice sites. The diamond is for the cooperative shielding that $|\psi(0)\rangle = 1/\sqrt{N}\sum_{n=1}^N |n\rangle$, the asterisk is for that $|\psi(0)\rangle = 1/\sqrt{5}\sum_{n=1}^5 |n\rangle$, and both of the two simulations are evaluated by right-y axis. (c) Comparison of cooperative shielding with $|c_n(T)|$ when $|\psi(0)\rangle = 1/\sqrt{N}\sum_{n=1}^N |n\rangle$ between $N=100$ (almost continuous trapping) and $N=10$ (inset, discrete trapping). (d) Comparison when $|\psi(0)\rangle = 1/\sqrt{5}\sum_{n=1}^5 |n\rangle$. In (b-d), $V=1$ and $U_0 = 0.1$.}
	\label{fig:AndercompareOnelayer}
\end{figure}

The trapped molecule array in Fig.~\ref{fig:Trapresult}(a) can be regarded as a lattice network with infinite number of sites and the spatial distribution of trapped localizations converges with the increase of $N$, as compared in Figs.~\ref{fig:AndercompareOnelayer}(a) and (b).

\subsection{Comparison of diffusion based on the single-layer trapped polar molecules}
Generalized from the case in Eq.~(\ref{con:pnevolution}), when the polar molecules are trapped by two SAW layers as shown in Fig.~\ref{fig:TraptopDesign}(a) without the external electric force, the molecules can be trapped at arbitrary locations along the $x$ direction, the spatial distribution of the molecules is governed by
\begin{small}
\begin{equation} \label{con:pnevolutionContinuous}
\begin{aligned}
\dot{p}_x(x,t) = &-iU_0p_x(x,t) - i\int_0^{x_-} v_{xx'} p_{x'}(x',t)\mathrm{d}x'\\
&- i\int_{x_+}^{M\lambda} v_{xx'} p_{x'}(x',t)\mathrm{d}x',
\end{aligned}
\end{equation}
\end{small}%
where $p_x(x,t)$ and $p_{x'}(x',t)$ are the probability amplitudes of the trapped molecules at $x$ and $x'$, respectively, $v_{xx'}\propto \left | x - x'\right|^{-3}$ represents the interaction between the sites at $x$ and $x'$, the parameters $M$ and $\lambda$ are the same as that in Eq.~(\ref{con:ExyvectorSimpleMain}).

The continuous model in Eq.~(\ref{con:pnevolutionContinuous}) can be approximated by the discrete model in Eq.~(\ref{con:pnevolution}) when $N\rightarrow \infty$ and $\Delta x = x_{n}-x_{n-1} \approx 0$ for $n =2,3,\cdots,N$. 
Then by Eqs.~(\ref{con:P1sSolution}) and (\ref{con:P1salphaBeta}), considering that the condition $U_0^3 - \alpha U_0 + \beta = 0$ can almost never be satisfied, we have $\lim_{t\rightarrow \infty}p_1(t)= \lim_{s\rightarrow 0}sP_1(s) = 0$. This agree with the case in Fig.~\ref{fig:AndercompareOnelayer}(a) when $N=500$.
\subsection{Cooperative shielding}
Together with the long range interactions, the evolution of the quantum states $|\psi(t) \rangle = \sum_{n=1}^N c_n(t) |n\rangle$ is governed by the initial state $|\psi(0)\rangle$ and the Hamiltonian~\cite{shieldingPRB,shieldingPRL}
\begin{footnotesize}
\begin{equation} \label{con:CoShieldHam}
\begin{aligned}
H_{N} = -V \sum_n \left ( |n\rangle \langle n+1| + H.C.\right ) - \sum_{n \neq j}\gamma_{nj} |n\rangle \langle j|-U_0 \sum_{n } |n\rangle \langle n|,
\end{aligned}
\end{equation}
\end{footnotesize}%
where $V$ is the amplitude for nearest-neighbour hopping, $\gamma_{nj}$ denotes long range interactions between the $n$th and $j$th lattices, and $U_0$ is the energy of the polar molecules on the site $n$. The influence by $\gamma_{nj}$ can be evaluated with the cooperative shielding effect~\cite{shieldingPRB,shieldingPRL,CopShieldPRB}, which can also be assessed by the occupation $c_n(T)$ at the terminal time point $T$.

We first consider a simplified case that $\gamma_{nj} \equiv \gamma$, and then consider a general case. As shown in Fig.~\ref{fig:AndercompareOnelayer}(b), where $c_n(T)$ is for $\gamma \neq 0$ and $c_n'(T)$ is for $\gamma = 0$~\cite{shieldingPRB}, the effect of $\gamma$ can be evaluated with $\max_{n}\left |\left |c_n(T) \right |^2 -\left |c_n'(T) \right |^2 \right |$. The comparisons in Figs.~\ref{fig:AndercompareOnelayer}(c) and (d) reveal that the long range interactions among trapped polar molecules can be cancelled out when the number of trapped sites is large based on the designed initial condition. 

According to the conclusion in Refs.~\cite{shieldingPRB,shieldingPRL}, we choose the initial quantum state as a random superposition. For the Hamiltonian given in Eq.~(\ref{con:CoShieldHam}), the quantum states with $N$ localized sites can be represented as
\begin{equation} \label{con:stateNsites}
\begin{aligned}
|\psi(t) \rangle = \sum_{n=1}^N c_n(t) |n\rangle,
\end{aligned}
\end{equation}%
where $c_n(t)$, as a generalization of $p_n$ in Eq.~(\ref{con:pnevolution}), represents the amplitude that the $n$th site is occupied. For the single-layer trapped polar molecules with a single SAW layer and externa electrical fields, $N$ can be finite. When the polar molecules are trapped with two SAW layers as in Fig.~\ref{fig:Trapresult}(a), $N$ can be infinitely large. 

\begin{figure}
\centerline{\includegraphics[width=0.9\columnwidth]{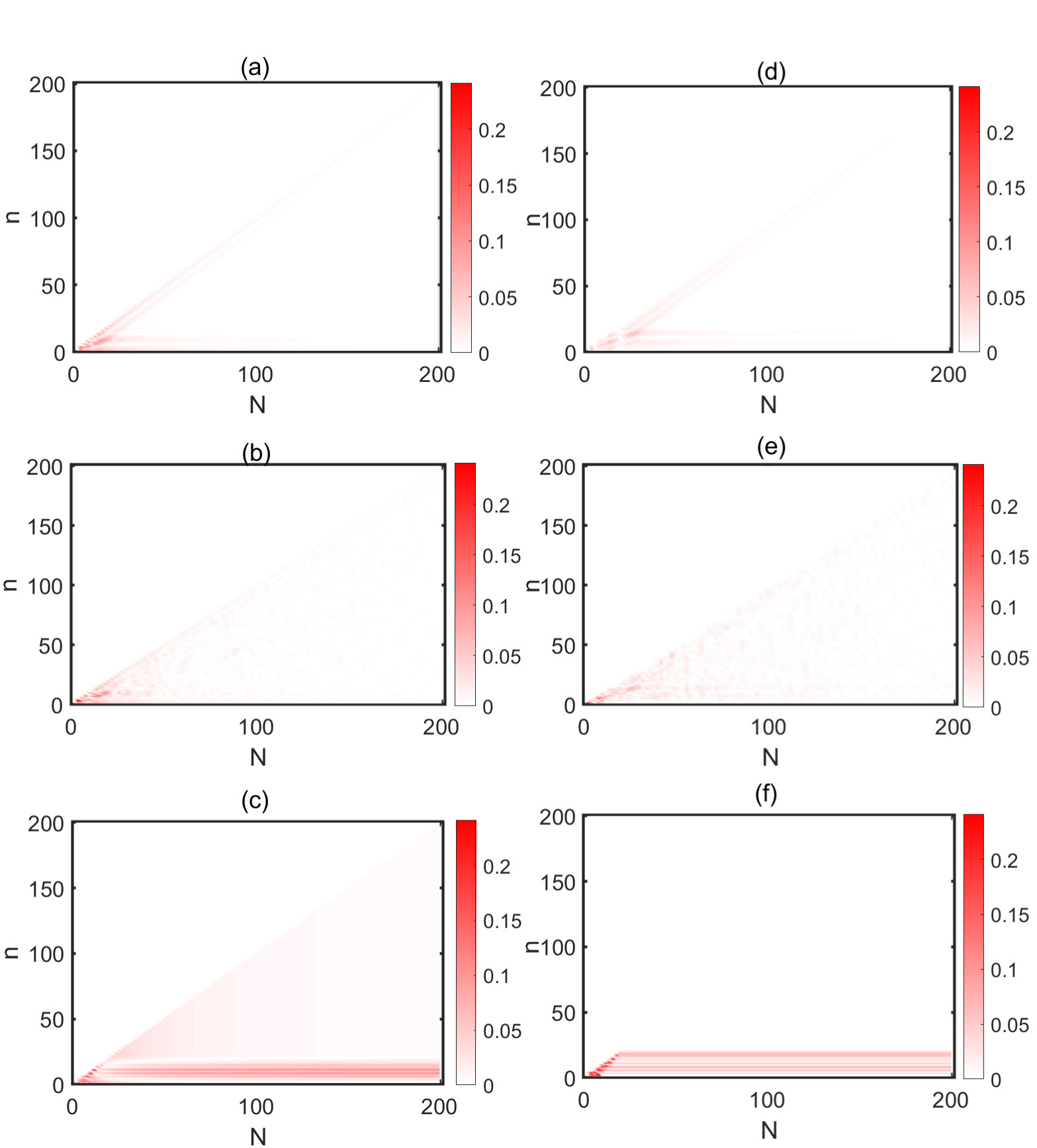}}
\caption{$\left|\left|c_n(T)\right|^2 - \left|c_n'(T)\right|^2\right|$ varies with the number of cites $N$, where $c_n(T)$ represents the terminal values in Eq.~(\ref{con:NsitesEvolution}) with $\gamma_{nj} = 0$ at $T = 10$s, and $c_n'(T)$ represents that when $\gamma_{nj} \neq0$. In all the simulations, $V=1$ and $U_0 = 0.1$. In (a-c), $\gamma_{nj} \equiv \gamma= 2$, and in (d-f), $\gamma_{nj} = V|n-j|^{-3}$. In (a) and (d), $c_n(0) = 1/\sqrt{N}$. In (b) and (e), $c_n(t)$ and $c_n'(t)$ are randomly initialized. In (c) and (f), $c_1(0) = c_2(0) = 1/\sqrt{2}$ and $c_n(0) = 0$ for $n>2$.}
	\label{fig:CooperativeShiN}
\end{figure}

Given the Hamiltonian as Eq.~(\ref{con:CoShieldHam}), 
the evolution of the amplitudes can be written according to the Schr\"{o}dinger equation $|\dot{\psi}(t) \rangle =-iH_N|\psi(t) \rangle$ as~\cite{shieldingPRB}
\begin{small}
\begin{equation} \label{con:NsitesEvolution}
\begin{aligned}
\dot{c}_n (t) &=iV c_{n+1}(t) + iV c_{n-1}(t) + i\sum_{n \neq j} \gamma_{nj} c_j(t) + iU_0  c_n(t),
\end{aligned}
\end{equation}
\end{small}%
with $\hbar  = 1$. In Fig.~\ref{fig:CooperativeShiN}, we compare the dynamics of Eq.~(\ref{con:NsitesEvolution}) with different number of sites and different initial conditions. The simulations clarify that long-range interactions can affect the dynamics of the lattices, but it is related to the initial condition and number of sites. If the lattices are initially random or uniformly occupied, the influence of long-range interactions is much smaller or better shielded than the case that a small subset of lattices are initially occupied.  

\section{Bose-Hubbard model for trapped molecules} \label{Sec:BHM}
When $\bar{u}_2\gg \bar{u}_1 > 0$, the polar molecules can be trapped in the single-layer lattices by $\bar{u}_2$ and external electrical force. This is similar to the purple dashed curve shown in Fig.~\ref{fig:Trapresult}(b) by replacing $z$ with $D-z$. The energy of the molecules is affected by electrical potential of the lower SAW, and the one-dimensional lattice of trapped polar molecules can be described by the Bose-Hubbard model with the Hamiltonian~\cite{BHMWannier,PRBExtended1,LatticeCiracZoller}
\begin{small}
\begin{equation} \label{con:HamEnergy}
\begin{aligned}
H &= \int \mathbf{\Psi}^{\dag} \left (\vec{r} \right ) \left [ -\frac{\hbar ^2 \nabla ^2 }{2m_0} + V_{e}\left (\vec{r} \right )\right ] \mathbf{\Psi}\left (\vec{r} \right ) \\
&+ \int \mathbf{\Psi}^{\dag} \left (\vec{r} \right )\mathbf{\Psi}^{\dag} \left (\vec{r'} \right ) V\left (\vec{r} - \vec{r'} \right )\mathbf{\Psi}\left (\vec{r'} \right )\mathbf{\Psi} \left (\vec{r} \right ) \mathrm{d}^3 \vec{r}  \mathrm{d}^3 \vec{r'},
\end{aligned}
\end{equation}
\end{small}%
where $m_0$ is the mass of the polar molecule, $V_e\left (\vec{r} \right )$ is the electrical potential by the lower SAW, and $\mathbf{\Psi}\left (\vec{r} \right )$ is a boson field operator for the trapped polar molecules. The field operator $\mathbf{\Psi}\left (\vec{r} \right )$ can be represented as a superposition of Wannier functions localized at the trapped lattice sites as
\begin{small}
\begin{equation} \label{con:FieldOperatorWannier}
\begin{aligned}
\mathbf{\Psi}\left (\vec{r} \right ) = \sum_{\mu}\hat{a}_{\mu} w\left (\vec{r} -\vec{r}_{\mu} \right ),
\end{aligned}
\end{equation}
\end{small}%
where the operator $\hat{a}_{\mu}$ annihilates a polar molecule at the $\mu$th site with $\vec{r}_{\mu}$, and $w$ is the Wannier function, which is determined by the distance between $\vec{r}$ and $\vec{r}_{\mu}$. 

When we only consider the nearest neighbor interaction between lattices, the trapped polar molecules can be described by the one-dimensional Bose-Hubbard model as~\cite{MottHubbardPRL,BHMPRL,PMLattice,SAWelectrongas2,NJPBHMPM,damski2003creation}
\begin{small}
\begin{equation} \label{con:BHM}
\begin{aligned}
H_{B} =&-J\sum_{\mu,\nu}\hat{a}_{\mu}^{\dag}\hat{a}_{\nu} + \frac{U}{2}\sum_{\mu}\hat{n}_{\mu}\left (\hat{n}_{\mu}-1\right ) + \sum_{\mu}\epsilon \hat{n}_{\mu},
\end{aligned}
\end{equation}
\end{small}%
where $\hat{a}_{\mu}\left (\hat{a}_{\mu}^{\dag} \right)$ annihilates (creates) a polar molecule at the $\mu$th site obeying the canonical commutation relation $\left [ \hat{a}_{\nu}, \hat{a}_{\mu}^{\dag} \right ] = \delta_{\nu\mu}$, $\hat{n}_{\mu} = \hat{a}_{\mu}^{\dag}\hat{a}_{\mu}$, $J$ represents the hopping amplitude between two lattice sites~\cite{BHMRMPReview}, $U$ is the on-site repulsion, and $\epsilon$ represents the energy offset of each lattice site affected by $\bar{u}_1$~\cite{SAWelectrongas2,LatticeCiracZoller,NJPBHMPM}.

The parameters in Eq.~(\ref{con:BHM}) are determined by different lattice sites $\mu$ and $\nu$, and can be given as~\cite{LatticeCiracZoller}
\begin{small}
\begin{equation} \label{con:JUcalculation}
   \begin{cases}
   J_{\mu\nu}&= \int w^*\left (\vec{r} -\vec{r}_{\mu} \right ) \left [ -\frac{\hbar ^2 \nabla ^2 }{2m_0} + V_{e}\left (\vec{r} \right )\right ] w\left (\vec{r} -\vec{r}_{\nu} \right ) \mathrm{d}^3 \vec{r},\\
   U&=\frac{4\pi}{m_0} \int | w\left(\vec{r} \right )|^4\mathrm{d}^3 \vec{r} ,\\
   \epsilon_{\mu} &=\int  V_{e}\left (\vec{r} \right ) \left | w\left(\vec{r} -\vec{r}_{\mu} \right )\right |^2\mathrm{d}^3 \vec{r},
   \end{cases}
\end{equation}
\end{small}%
where $V_{e}\left (\vec{r} \right )$ is the electrical potential induced by the trapping IDT as in Eqs.~(\ref{con:Potential0}) and (\ref{con:TotalPotential}) in Appendix~\ref{sec:calWave1}. The Wannier function can be approximately simplified with the Gaussian function as $w(x) = e^{-x^2/2}$~\cite{WannierAppro,WannierAppro2,WannierAppro3}. When $|\mu-\nu| = 1$ in Eq.~(\ref{con:JUcalculation}), $J_{\mu\nu}$ reduces to $J$ for the simplified nearest neighbor case in Eq.~(\ref{con:BHM}). 

Above all, the lattice network is determined by the spatial distribution of the external force, and the number of lattice sites can be arbitrarily manipulated. If we only consider a short time scale that $k(x-vt)$ varies a little, as the practical case in Refs.~\cite{meek2009trapping,meekPRL,meekNJP,meekphdthesis}, or the theoretical estimation in Ref.~\cite{SAWelectrongas2} that the tunneling time among lattices and the time for the occurrence of lattice dynamics vary from $10~\rm ps$ to $15~\rm ns$, then Eq.~(\ref{con:JUcalculation}) can be simplified to
\begin{small}
\begin{equation} \label{con:Jmunu}
\begin{aligned}
 J_{\mu\nu}
 &= \int e^{-\left (x - x_{\mu}\right )^2/2}  \left \{ -\frac{\hbar ^2 \nabla ^2 }{2m_0} \right.\\
&~~~~\left. + B_0 e^{-kz} \cos \left [k(x-vt) \right ] \right \}  e^{-\left (x - x_{\nu}\right )^2/2} \mathrm{d}x \\
&\approx \Delta \left (\frac{1-\Delta^2}{2m_0} + B_0 e^{-kz} \cos \left [k(x-vt) \right ]\right )e^{-\Delta^2}\\
  &\approx \Delta \left ( \tilde{B}_0 e^{-kz} + \delta_J \right )e^{-\Delta^2}\\
 &\triangleq J,
\end{aligned}
\end{equation}
\end{small}%
where $\Delta$ is the distance between two nearest neighbor lattice sites, $\hbar =  1$ for simplification, then we regard the time-domain evolution as a bounded perturbation as $\delta_J$~\cite{BHMRMPReview}. Similarly,
\begin{small}
\begin{equation} \label{con:JUcalculation2}
   \begin{cases}
   U&=\frac{4\pi}{m_0} \int e^{-2x^2} \mathrm{d}x ,\\
   \epsilon &= B_0 e^{-kz} \cos \left [k(x_{\nu}-vt) \right ] \approx \left ( \tilde{B}_0 + \delta_{\epsilon}\right) e^{-kz},
   \end{cases}
  \end{equation}
\end{small}%
where $\delta_{\epsilon}$ is a bounded perturbation similarly as in Eq.~(\ref{con:Jmunu}) within a short time scale.
In the following, we first study a simplified case with $ \delta_J=\delta_{\epsilon}=0$ and then generalize to the case that $ \delta_J , \delta_{\epsilon} \neq 0$. 

\begin{figure}[h]
\centerline{\includegraphics[width=1\columnwidth]{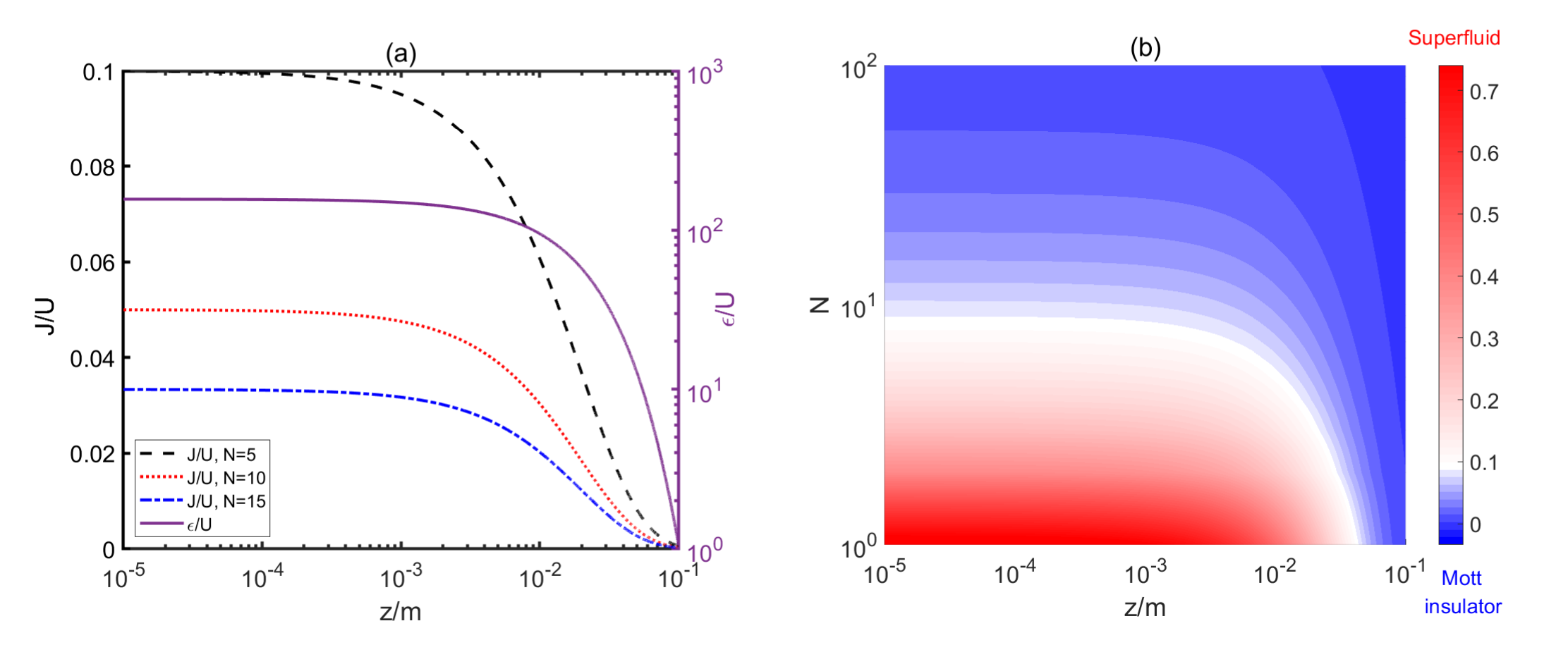}}
\caption{
(a) $J/U$ and $\epsilon/U$ affected by the trapped height $z$ of polar molecules. The distance between two nearest neighbor sites is $\Delta = 0.005/N$m with $N=5,10,15$ representing the number of lattice sites, $ \delta_J = \delta_{\epsilon} = 0$, $m_0 = 0.1$, $\tilde{B}_0 = 100$ and $k =50$.
(b) The phase transition evaluated by $J/U$ between the Mott insulator and superfluid affected by the trapped height and the number of lattice sites. In the simulation, $k=50$, the polar molecules are trapped by the IDT array with a horizontal width of $0.005$m, $n_0 = 1$ and $f_{n_0} = 0.085786$.}
	\label{fig:compareOnelayer}
\end{figure}

In Fig.~\ref{fig:compareOnelayer}, $J/U$ and $\epsilon/U$ are plotted as functions of $z$, which is the distance between the trapped polar molecules and the lower SAW. It can be seen that both $J/U$ and $\epsilon/U$ decrease with the increase of $z$. This is because of the fact that the electrical potential by the lower SAW decreases with the increase of $z$. Using this, we can modulate the transition between different phases of the Bose-Hubbard model as follows.

The transition between the superfluid and Mott insulator is determined by $J/U$ and the average number of molecules $n_0$ at each lattice site~\cite{BHMPRBtransition}. According to the conclusion in Refs.~\cite{BHMPRBtransition,fisher1989boson}, the transition boundary between the superfluid and Mott insulator occurs when $\epsilon/U -n_0 = -1/2 -J/U \pm \sqrt{(J/U)^2-\left ( 2n_0+1\right)J/U + 1/4}$, then the polar molecule gas will be superfluid when $J/U > n_0+1/2 - \sqrt{n_0\left(n_0 + 1\right)} \triangleq f_{n_0}$, where $n_0=1,2,\cdots$ represent the mean polar molecule number in the trapped lattice sites. For example, when $n_0 = 1$, $f_{n_0} = 0.085786$, corresponds to the case in Fig.~\ref{fig:compareOnelayer}(b). 
The relationship between the designed $J/U$ and $\epsilon$ can affect the transition between the superfluid ($J/U > f_{n_0}$) and Mott insulator ($J/U <f_{n_0} $)~\cite{NJPBHMPM,PhaseFomular,freericks1994phase,kuhner1998phases}. Thus by controlling the trapped position of polar molecules along $z$ direction and the number of lattice sites, the transition between the superfluid and Mott insulator can be manipulated as in Fig.~\ref{fig:compareOnelayer}(b). In this transition, the exponential decay of the electrical potential induced by SAW in the open space along the $z$-direction plays a crucial role, and makes it possible for the transition occurrence with varied numbers of lattice sites, which is the special property provided by the acoustic wave when compared with traditional modulating methods~\cite{YinJPThreeD,Meijerdeceleration,ElectrostaticTrapVelocity,TwofieldTrap}.

Additionally, the transition process is also affected by $n_0$ and the uncertainties induced by time dependent potential if we only consider a quite short time period. In such cases, the molecules are trapped in the lattices as in experimental circumstance in Ref.~\cite{meek2009trapping}, and are also affected by the thermal effect determined by the environment temperature. The phase transition needs to be further analyzed as follows.  

\subsection{Influence by $n_0$, uncertainties and thermal effects}
\begin{figure}[h]
\centerline{\includegraphics[width=1\columnwidth]{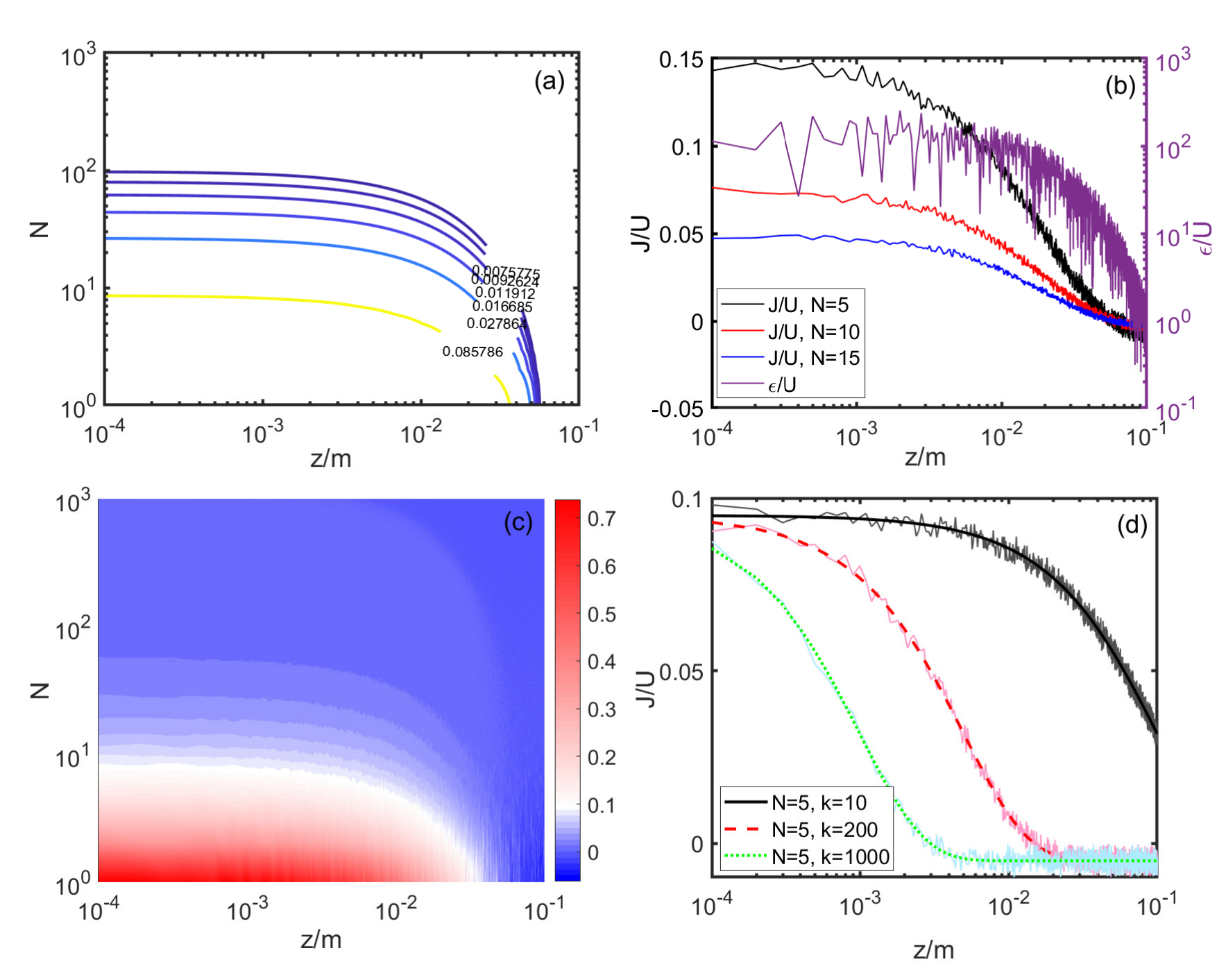}}
\caption{(a) Phase transition boundaries when $n_0 = 1,4,7,10,13,16$ and $f_{n_0} = 0.085786, 0.027864,\cdots,0.0075775$, respectively. (b) The parameters of the Bose-Hubbard model affected by random components $\delta_J\in [-5,5]$ and $\delta_{\epsilon}\in [-100,100]$. (c) The phase transition between superfluid (red) and Mott insulator (blue) evaluated by $J/U$ with the uncertain parameters in (b). (d) $J/U$ affected by the wave number $k$ when $N=5$ and $\delta_J\in [-5,5]$. In all the simulations, $\tilde{B}_0=100$, the distance between two nearest neighbor trapped polar molecules is $0.005/N$m, and $k = 50$ in (a)-(c).}
	\label{fig:MottTransition}
\end{figure}
For a general case of $n_0$, the contours in Fig.~\ref{fig:MottTransition}(a) represent the phase transition boundary when $n_0 = 1,4,7,10,13,16$, where $f_{n_0}$ decreases with the increase of $n_0$. Thus when $n_0$ is larger, it is easier to generate the superfluid state. Fig.~\ref{fig:MottTransition}(b) represents the parameters with random oscillations, 
and Fig.~\ref{fig:MottTransition}(c) represents the phase transition affected by the parameters in Fig.~\ref{fig:MottTransition}(b) when $n_0=1$. It can be seen that the parameter oscillations can induce some errors but not destruct the overall transition property mainly affected by the trapped height of the polar molecule array and the number of lattice sites.

\begin{figure}[h]
\centerline{\includegraphics[width=1\columnwidth]{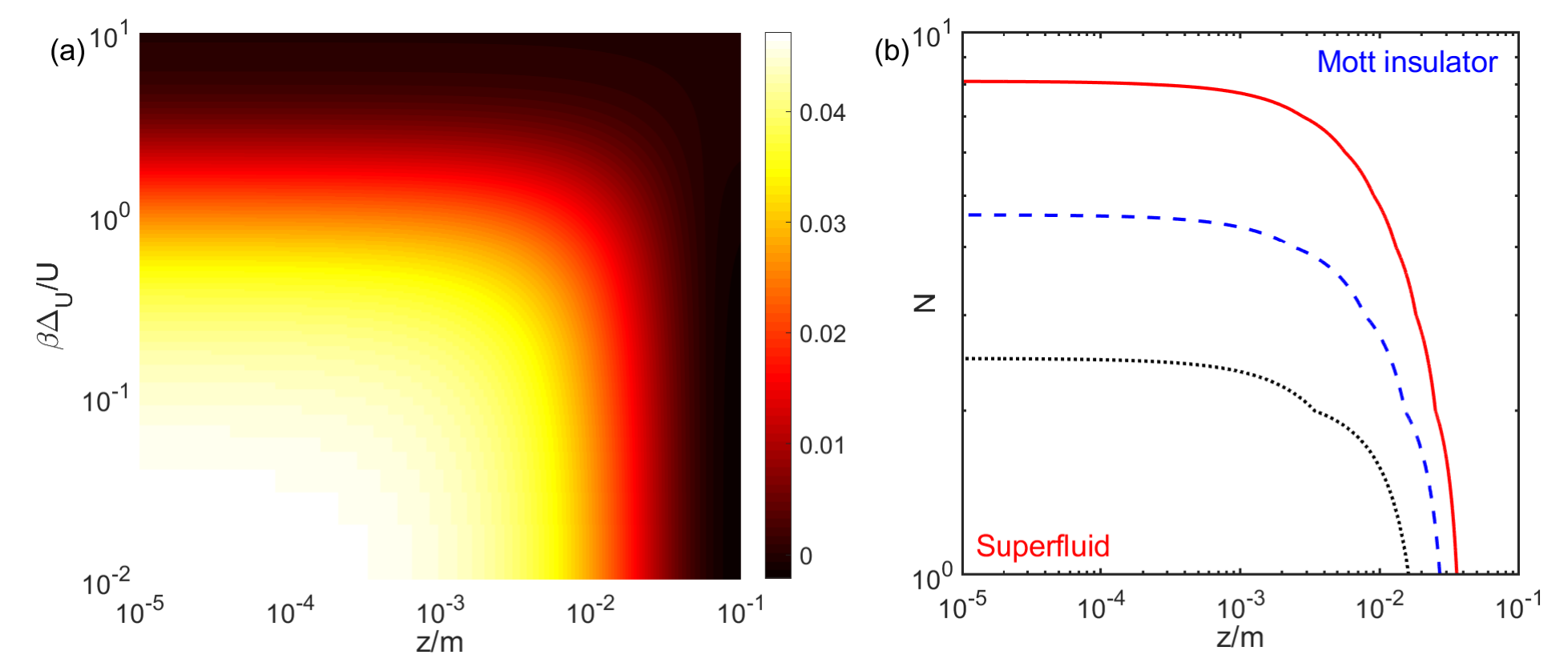}}
\caption{(a) $J_T/U$ affected by the temperature when $N=10$ and other parameters are same as Fig.~\ref{fig:compareOnelayer}(a). (b) The phase transition boundary between superfluid and Mott-insulator for $n_0 =1$ affected by different temperatures with $\beta \Delta_U = 0.1U$(solid-red), $U$(dash-blue) and $2U$(dot-black), respectively.}
	\label{fig:Temperature}
\end{figure}

Besides, the thermal effects can affect the hopping of polar molecules among different lattice sites, and this will further affect the parameter $J$ in the Bose-Hubbard model in Eq.~(\ref{con:BHM}) as $J_T = J e^{-\beta \Delta_U}$~\cite{stasyuk2009phase,mahan1976lattice,stasyuk2007density}, where $\beta$ is the temperature, $\Delta_U$ is a constant and $\beta \Delta_U$ can be evaluated according to $U$ in the Bose-Hubbard model. Then the evolutions of lattices are governed by Eq.~(\ref{con:BHM}) after replacing $J$ with $J_T$. As illustrated in Fig.~\ref{fig:Temperature}(a), the thermal effect can affect the parameter $J_T/U$, together with the trapped height, and this can further affect the phase transitions of the trapped polar molecule gas. As compared in Fig.~\ref{fig:Temperature}(b), the lower temperature is easier for the construction of superfluid state, and stronger thermal effect can induce Mott insulator. 

\subsection{Influence by the wave number}
\begin{figure*}
\centerline{\includegraphics[width=2\columnwidth]{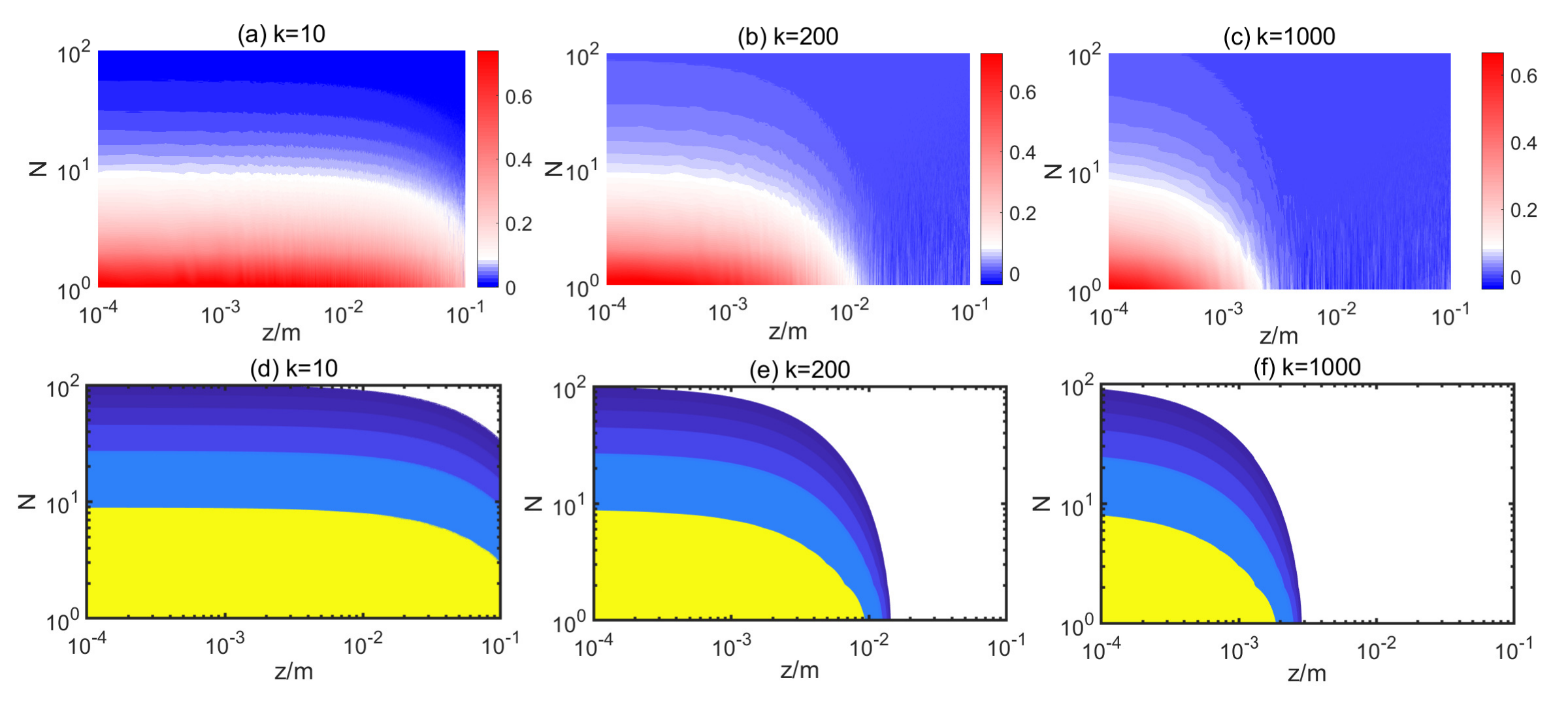}}
\caption{The influence on the phase transition by the wave number $k$ of the acoustic wave. (a)-(c) represent the phase transition between the superfluid (red region) and Mott insulator (blue region) with $n_0 = 1$ and the uncertainties in Fig.~\ref{fig:MottTransition} (b). (d)-(f) represent the phase transition boundaries with $n_0 = 1,4,7,10,13,16$, and different colors represent the regions for the superfluid state separated by the boundaries with different $n_0$ compared with the white insulator region. In all the simulations, $B_0 = 100$, the distance between two nearest neighbor trapped polar molecules is $0.005/N$m.}
	\label{fig:TransitionTestk}
\end{figure*}

For different wave numbers of the acoustic wave, the phase transition boundaries are different. As shown in Eq.~(\ref{con:ExyvectorSimple}) and Fig.~\ref{fig:MottTransition}(d), the amplitudes of the electric field and the electrical force by SAW decrease along the $z$ direction with the rate determined by the wave number $k$. When $k$ is small, the transition between the Mott insulator and superfluid is mainly determined by the number of lattice sites, as shown in Figs.~\ref{fig:TransitionTestk} (a) and (d). In this case, when $N$ is small, the molecule array can mainly be the superfluid if the molecules are only trapped close to the electrodes on the piezoelectric material with surface acoustic waves. However, when $k$ is large, the electrical potential decreases fast with the increase of $z$. Thus when $N$ is small, the transition between the superfluid and Mott insulator can also occur when $k$ is large enough, as shown in Figs.~\ref{fig:TransitionTestk}(b-f).

\section{Conclusion} \label{Sec:Conclusion}
We have proposed a method to trap polar molecules with the electrical field induced by surface acoustic waves. Assisted by external electrical fields, the polar molecules can be trapped and arranged into single or multi-parallel layers away from the piezoelectric material. Depending on the uniformity of the layers in the longitudinal direction, the attractive interactions among the molecules in each layer can be different, which can further affect the binding energy of the trapped polar molecule array. For a single layer of trapped polar molecules, the final steady distribution of the molecules can be affected by the trapping approach with finite or infinite lattice sites. The trapped polar molecules can be used to construct the lattice based Bose-Hubbard model. The phase transition between the superfluid and Mott insulator can be controlled by designing the values and spatial distributions of the external electrical force.
The advantages of SAW for trapping polar molecules are that its induced electrical field satisfies the condition of stability for trapping a polar particle, the field can be real-time modulated by the electrical field applied to the input IDT, and arbitrary three-dimensional polar molecule lattices can in principle be trapped by tuning another applied electrical force. Our proposal can be more efficient compared with traditional trapping methods with electrical force such as in Refs.~\cite{YinJPThreeD,Meijerdeceleration,ElectrostaticTrapVelocity,TwofieldTrap}. Besides, the exponential decrease of the electrical field produced by SAW along the vertical direction in the open space can induce the transition between Mott insulator and superfluid, and the overall architecture can be fabricated on-chip with extensive possible expansions.
From the perspective of experimental realization, the deceleration of the molecule gas has been realized in Refs.~\cite{meek2009trapping,meekPRL,quintero2013static}, the on-chip design of the surface acoustic wave resonator has been demonstrated in Ref.~\cite{AlekseyPRL}, and the method of shaping electrical fields in the open space between two SAW layers has been introduced in Refs.~\cite{YinJPThreeD,Yin2022driven}. These technologies support the viability of the proposed trapping method. Our proposal has potential applications in the construction of superfluid and Mott insulator~\cite{MottPM,ZhangJingSXNature}, phase transitions~\cite{PhasePM,wang2007quantum,greiner2002quantum}, lattice QED~\cite{PMtopoThreeDim,PMLattice}, as well as quantum information processing~\cite{Jiang2019QRAM}. This might open an easy way to hybridize molecules with solid state quantum devices, in contrast to the trapping molecules using optical lattices.

\emph{Acknowledgement.---}
Y.X.L. is supported by the National Science Foundation of China (NSFC) under Grants No. 12374483 and No. 92365209. R.B.W. is supported by the NSFC under Grants No. 61833010 and
No. 62173201.

\appendix

\section{Calculations on the surface acoustic wave}  \label{sec:calWave1}
Assume the displacement of the piezoelectric material is $\mathbf{u} = (u_1,u_2,u_3)$ at the position $\vec{r}= [x_1,x_2,x_3]^T$. When there are not any strains in the material, the material moves as a whole and the displacement at $\vec{r}$ and $\vec{r'}$ satisfies that
\begin{small}
\begin{eqnarray} \label{eq:strain0}
\mathbf{u}(\vec{r}) = \mathbf{u}(\vec{r'}),
\end{eqnarray}
\end{small}%
and
\begin{small}
\begin{eqnarray} \label{eq:strain1}
\frac{\partial u_i}{\partial x_j} = 0, ~~~~\forall i,j = 1,2,3.
\end{eqnarray}
\end{small}

The strain at $\vec{r}$ is defined as
\begin{small}
\begin{equation} \label{con:strain}
\begin{aligned}
S_{ij}(x_1,x_2,x_3)=\frac{1}{2}\left(\frac{\partial u_i}{\partial x_j}+\frac{\partial u_j}{\partial x_i}\right),~i,j=1,2,3.
\end{aligned}
\end{equation}
\end{small}

If the material is not piezoelectric, the motion of a single unit satisfies that~\cite{morgan2010surface}
\begin{small}
\begin{equation} \label{con:NewtonMotion}
\begin{aligned}
\rho \frac{\partial^2 u_i}{\partial t^2} = \sum_j \frac{\partial T_{ij}}{\partial x_j}, ~i,j = 1,2,3,
\end{aligned}
\end{equation}
\end{small}%
where $\rho$ is the density, $T$ is the stress. Here $T$ is proportional to the strain $S$ if the inner force is small, namely
\begin{small}
\begin{equation} \label{con:TS}
\begin{aligned}
T_{ij} = \sum_k\sum_l c_{ijkl}S_{kl},~i,j,k,l =1,2,3,
\end{aligned}
\end{equation}
\end{small}%
where $S_{kl} = \partial u_k/\partial x_l$, the parameters $c_{ijkl}$ are regarded as the stiffness tensor and can be replaced with the matrix $\mathbf{C}$ as $$\mathbf{C}=\begin{bmatrix}
c_{11} & c_{12} & c_{12} & 0 & 0 & 0\\
c_{12} & c_{11} & c_{12} & 0 & 0 & 0\\
c_{12} & c_{12} & c_{11} & 0 & 0 & 0\\
0 & 0 & 0 & c_{44} & 0 & 0\\
0 & 0 & 0 & 0 & c_{44} & 0\\
0 & 0 & 0 & 0 & 0 & c_{44}\\
\end{bmatrix}$$ with the corresponding relationship between $c_{ijkl}$ and $\mathbf{C}$ introduced in Ref.~\cite{morgan2010surface}.

\subsection{Solutions in isotropic materials}
In the isotropic material,
\begin{small}
\begin{equation} \label{con:cijkl}
\begin{aligned}
c_{ijkl} = \kappa \delta_{ij}\delta_{kl} + \mu\left(\delta_{ik}\delta_{jl} + \delta_{il}\delta_{jk}\right),
\end{aligned}
\end{equation}
\end{small}%
where $\kappa$ and $\mu$ are positive Lam$\acute{e}$ constants and the Kronecker delta function is defined as $\delta_{ij}= 1$ when $i=j$ and $\delta_{ij}= 0$ when $i\neq j$. The stress $T$ can be calculated as
\begin{small}
\begin{equation} \label{con:TSE_isotropic}
\begin{aligned}
T_{ij} &= \sum_k\sum_l c_{ijkl}S_{kl}\\
& = \sum_k\sum_l \left(\kappa \delta_{ij}\delta_{kl} + \mu\left(\delta_{ik}\delta_{jl} + \delta_{il}\delta_{jk}\right)\right)S_{kl}  \\
& = \sum_k\sum_l \kappa \delta_{ij}\delta_{kl} S_{kl} +  \sum_k\sum_l \mu\left(\delta_{ik}\delta_{jl} + \delta_{il}\delta_{jk}\right) S_{kl} \\
& = \sum_k \kappa \delta_{ij} S_{kk} +  \sum_k\sum_l \mu\left(\delta_{ik}\delta_{jl} + \delta_{il}\delta_{jk}\right) S_{kl} \\
& = \sum_k \kappa \delta_{ij} S_{kk} + 2 \mu S_{ij},
\end{aligned}
\end{equation}
\end{small}%
where
\begin{small}
\begin{equation} \label{con:Skk}
\begin{aligned}
S_{kk} = \frac{\partial u_k}{\partial x_k}.
\end{aligned}
\end{equation}
\end{small}%
Denote $\Delta = \sum_k S_{kk}$, then
\begin{small}
\begin{equation} \label{con:TSE_isotropic2}
\begin{aligned}
T_{ij} &= \kappa \delta_{ij} \Delta + 2 \mu S_{ij}.
\end{aligned}
\end{equation}
\end{small}%
Equation~(\ref{con:NewtonMotion}) can be written as
\begin{small}
\begin{equation} \label{con:NewtonMotion2}
\begin{aligned}
\rho \frac{\partial^2 u_i}{\partial t^2} &= \sum_j \frac{\partial T_{ij}}{\partial x_j}\\
&= \frac{\partial \kappa \Delta}{\partial x_i} + 2 \mu\sum_j \frac{\partial S_{ij}}{\partial x_j} \\
&=  \kappa \frac{\partial \Delta}{\partial x_i} + 2 \mu \sum_j  \frac{\frac{1}{2}\left(\frac{\partial u_i}{\partial x_j}+\frac{\partial u_j}{\partial x_i}\right)}{\partial x_j} \\
&= \kappa \frac{\partial \Delta}{\partial x_i} +  \mu \sum_j \frac{\partial^2 u_j}{\partial x_i \partial x_j}+ \mu \sum_j \frac{\partial^2 u_i}{\partial x_j^2} \\
&= \kappa \frac{\partial \Delta}{\partial x_i} + \mu \frac{\partial \left(\sum_j \partial u_j/\partial x_j\right)}{\partial x_i}  + \mu \nabla^2 u_i \\
&=\left(\kappa + \mu\right)\frac{\partial \Delta}{\partial x_i} + \mu \nabla^2 u_i,
\end{aligned}
\end{equation}
\end{small}%
where $\nabla^2 = \sum_i \frac{\partial^2}{\partial x_i^2}$ and $i,j = 1,2,3$.

The components of Eq.~(\ref{con:NewtonMotion2}) read
\begin{small}
\begin{equation} \label{con:NewtonMotionx1}
\begin{aligned}
&~~~~\rho \frac{\partial^2 u_1}{\partial t^2} =(\kappa + \mu)\frac{\partial (S_{11} + S_{22} + S_{33})}{\partial x_1} + \mu \nabla^2 u_1 \\
& = (\kappa + 2\mu)\frac{\partial^2 u_1}{\partial^2 x_1} + (\kappa + \mu)\left(\frac{\partial^2 u_2}{\partial x_2 \partial x_1} + \frac{\partial^2 u_3}{\partial x_3 \partial x_1}\right) \\
&~~~~+ \mu \left(\frac{\partial^2 u_1}{\partial^2 x_2} +\frac{\partial^2 u_1}{\partial^2 x_3}\right),
\end{aligned}
\end{equation}
\end{small}
\begin{small}
\begin{equation} \label{con:NewtonMotionx2}
\begin{aligned}
\rho \frac{\partial^2 u_2}{\partial t^2} & = (\kappa + 2\mu)\frac{\partial^2 u_2}{\partial^2 x_2} + (\kappa+ \mu)\left(\frac{\partial^2 u_1}{\partial x_2 \partial x_1} + \frac{\partial^2 u_3}{\partial x_3 \partial x_2}\right) \\
&~~~~+ \mu \left(\frac{\partial^2 u_2}{\partial^2 x_1} +\frac{\partial^2 u_2}{\partial^2 x_3}\right),
\end{aligned}
\end{equation}
\end{small}%
and
\begin{small}
\begin{equation} \label{con:NewtonMotionx3}
\begin{aligned}
\rho \frac{\partial^2 u_3}{\partial t^2} & = (\kappa + 2\mu)\frac{\partial^2 u_3}{\partial^2 x_3} + (\kappa + \mu)\left(\frac{\partial^2 u_1}{\partial x_3 \partial x_1} + \frac{\partial^2 u_2}{\partial x_3 \partial x_2}\right)\\
&~~~~ + \mu \left(\frac{\partial^2 u_3}{\partial^2 x_1} +\frac{\partial^2 u_3}{\partial^2 x_2}\right).\\
\end{aligned}
\end{equation}
\end{small}

The vector format of the displacement $\mathbf{u} = \left[u_1,u_2,u_3\right]^T$ can be represented as~\cite{stoneley1955}
\begin{small}
\begin{equation} \label{con:displacement}
\begin{aligned}
\left[u_1,u_2,u_3\right]^T = \left[\bar{U},\bar{V},\bar{W}\right]^Te^{-kqx_3+ik(lx_1+mx_2-vt)},
\end{aligned}
\end{equation}
\end{small}%
where $\bar{U}$, $\bar{V}$ and $\bar{W}$ are constants determined by $q$ representing the possible decay in the direction of $x_3$, $k$ is the wave number, $v$ is the velocity of the acoustic wave, $l = \cos\theta$, $m = \sin\theta$, $\theta$ is the angle between the propagating direction of the wave front with the $x_3$-axis. Thus $l$ and $m$ can represent the propagating direction of the acoustic wave. Take Eq.~(\ref{con:displacement}) into Eqs.~(\ref{con:NewtonMotionx1},\ref{con:NewtonMotionx2},\ref{con:NewtonMotionx3}), we have  $\mathbf{P}(q)\left[\bar{U}~\bar{V}~i\bar{W}\right]^T = 0$
with the matrix~\cite{CiracPRX} 
\begin{small}
\begin{equation}
\mathbf{P}(q)=\begin{bmatrix}
P_{11} & (\kappa + \mu) l m & (\kappa + \mu)lq \\
(\kappa + \mu)l m & P_{22} & (\kappa + \mu)mq \\
(\kappa + \mu) l q & (\kappa + \mu)mq & P_{33}  \\
\end{bmatrix},
\end{equation}
\end{small}%
where $P_{11} = (\kappa + 2\mu)l^2 -v^2 \rho + \mu (m^2 - q^2)$, $P_{22} = (\kappa + 2\mu) m^2 -v^2 \rho  + \mu(l^2 -q^2)$, $P_{33} = (\kappa + 2\mu) q^2 +v^2 \rho -\mu(l^2 + m^2)$, $l^2 + m^2 = 1$. 
Take $\det(\mathbf{P}(q)) = 0$, and we can denote the solutions of $q$ as $q_1,q_2,q_3$.

\subsection{Solutions in piezoelectric materials}
If the material is piezoelectric, the stress is also determined by the external electrical field, namely~\cite{morgan2010surface}
\begin{small}
\begin{equation} \label{con:TSE}
\begin{aligned}
T_{ij} = \sum_k\sum_l c_{ijkl}S_{kl} - \sum_k e_{kij}E_k,~i,j,k,l =1,2,3,
\end{aligned}
\end{equation}
\end{small}%
where the electrical field $E_k$ is applied along the x-axis.
Assume the electrical potential is $\Phi$, the electrical field satisfies that $E_k = -\partial \Phi/\partial x_k$. Thus the motion equation within the piezoelectric materials reads
\begin{small}
\begin{equation} \label{con:TSEMotionEquation}
\begin{aligned}
&~~~~\rho \frac{\partial^2 u_i}{\partial t^2} = \sum_j \frac{\partial T_{ij}}{\partial x_j}\\
& = \sum_j \frac{\partial \left(\sum_k\sum_l c_{ijkl}S_{kl} - \sum_k e_{kij}E_k\right)}{\partial x_j} \\
& = \sum_j \sum_k e_{kij} \frac{\partial^2 \Phi}{\partial x_j \partial x_k} + \sum_j \sum_k \sum_lc_{ijkl}\frac{\partial S_{kl}}{\partial x_j} \\
& = \sum_j \sum_k e_{kij} \frac{\partial^2 \Phi}{\partial x_j \partial x_k} + \frac{1}{2}\sum_j \sum_k \sum_lc_{ijkl}\frac{\partial \left(\frac{\partial u_k}{\partial x_l}+\frac{\partial u_l}{\partial x_k}\right)}{\partial x_j} \\
& = \sum_j \sum_k e_{kij} \frac{\partial^2 \Phi}{\partial x_j \partial x_k} + \frac{1}{2}\sum_j \sum_k \sum_lc_{ijkl}\left(\frac{\partial^2 u_k}{\partial x_l\partial x_j} + \frac{\partial^2 u_l}{\partial x_k\partial x_j}\right) \\
& = \sum_j \sum_k e_{kij} \frac{\partial^2 \Phi}{\partial x_j \partial x_k} + \frac{1}{2}\sum_j \sum_k \sum_lc_{ijkl}\frac{\partial^2 u_k}{\partial x_l\partial x_j}\\
&~~~~ +  \frac{1}{2}\sum_j \sum_k \sum_lc_{ijkl}\frac{\partial^2 u_l}{\partial x_k\partial x_j}.
\end{aligned}
\end{equation}
\end{small}%
Because $S_{kl} = S_{lk}$ and $T_{ij} = T_{ji}$, then
\begin{small}
\begin{equation} \label{con:TSEMotionEquation2}
\begin{aligned}
\frac{\partial^2 u_k}{\partial x_l\partial x_j} = \frac{\partial^2 u_l}{\partial x_k\partial x_j},
\end{aligned}
\end{equation}
\end{small}%
and
\begin{small}
\begin{equation} \label{con:TSEMotionEquation2}
\begin{aligned}
\rho \frac{\partial^2 u_i}{\partial t^2} & = \sum_j \sum_k e_{kij} \frac{\partial^2 \Phi}{\partial x_j \partial x_k} + \sum_j \sum_k \sum_lc_{ijkl}\frac{\partial^2 u_k}{\partial x_l\partial x_j}. \\
\end{aligned}
\end{equation}
\end{small}

The piezoelectric material is considered to be an insulator with $\int_v \nabla \cdot \mathbf{D} dv= 0$, and $\mathbf{D} = \left[D_i, D_j, D_k\right ]$ with $D_i = D_j = 0$, then
\begin{small}
\begin{equation} \label{con:divD}
\begin{aligned}
div \mathbf{D} &= \frac{\partial D_i}{\partial x_i} + \frac{\partial D_j}{\partial x_j} + \frac{\partial D_k}{\partial x_k}\\
&=-\sum_j\varepsilon_{kj}^S\frac{\partial^2 \Phi}{\partial x_j\partial x_k} + \frac{1}{2}\sum_i\sum_je_{ijk}\left(\frac{\partial^2 u_i}{\partial x_j\partial x_k}+\frac{\partial^2 u_j}{\partial x_i\partial x_k}\right),
\end{aligned}
\end{equation}
\end{small}%
where
\begin{small}
\begin{equation} \label{con:Dk}
\begin{aligned}
D_k &= \sum_j\varepsilon_{kj}^SE_j + \sum_i\sum_je_{ijk}S_{ij} \\
&= -\sum_j\varepsilon_{kj}^S\frac{\partial \Phi}{\partial x_j} + \frac{1}{2}\sum_i\sum_je_{ijk}\left(\frac{\partial u_i}{\partial x_j}+\frac{\partial u_j}{\partial x_i}\right).
\end{aligned}
\end{equation}
\end{small}

Because $i$ and $j$ are in the symmetric positions as $\partial^2 u_i/\partial x_j/\partial x_k = \partial^2 u_j/\partial x_i/\partial x_k$, then
\begin{small}
\begin{equation} \label{con:divD2}
\begin{aligned}
div \mathbf{D} &=-\sum_j\varepsilon_{kj}^S\frac{\partial^2 \Phi}{\partial x_j\partial x_k} + \sum_i\sum_je_{ijk}\frac{\partial^2 u_i}{\partial x_j\partial x_k}.
\end{aligned}
\end{equation}
\end{small}%
Then
\begin{small}
\begin{equation} \label{con:divD3}
\begin{aligned}
\sum_k \left(\sum_i\sum_je_{ijk}\frac{\partial^2 u_i}{\partial x_j\partial x_k} -\sum_j\varepsilon_{kj}^S\frac{\partial^2 \Phi}{\partial x_j\partial x_k}\right) = 0 .
\end{aligned}
\end{equation}
\end{small}

Because of the piezoelectric property, the boundary condition reads,
\begin{small}
\begin{equation} \label{con:Boundary}
\begin{aligned}
T_{iz} &= \sum_k\sum_l c_{izkl}S_{kl} - \sum_k e_{kiz}E_k\\
&= \sum_k\sum_l c_{izkl}S_{kl} + \sum_k e_{kiz} \frac{\partial \Phi}{\partial x_k}  \\
&= \sum_k\sum_l c_{izkl}\frac{1}{2}\left(\frac{\partial u_k}{\partial x_l}+\frac{\partial u_l}{\partial x_k}\right) + \sum_k e_{kiz} \frac{\partial \Phi}{\partial x_k} = 0,
\end{aligned}
\end{equation}
\end{small}%
where $i,k,l =1,2,3$ and $z=0$.

Because $S_{kl} = S_{lk}$, then
\begin{small}
\begin{equation} \label{con:Boundary2}
\begin{aligned}
T_{iz} = \sum_k\sum_l c_{izkl}\frac{\partial u_k}{\partial x_l} + \sum_k e_{kiz} \frac{\partial \Phi}{\partial x_k} = 0, \\
\end{aligned}
\end{equation}
\end{small}%
which means that there are no mechanical forces in arbitrary directions on the free surface, namely $T_{zx} = T_{zy} = T_{zz} = 0$ at $z = 0$. Then Eq.~(\ref{con:Boundary2}) can be rewritten as
\begin{small}
\begin{equation} \label{con:Boundary3}
\begin{aligned}
c_{izkl}\frac{\partial u_k}{\partial x_l} + e_{kiz}\frac{\partial \Phi}{\partial x_k} = 0,~\forall i,k,l =1,2,3. \\
\end{aligned}
\end{equation}
\end{small}%
Above all, the equations of the acoustic field can be written as
\begin{small}
\begin{eqnarray}  \label{eq:AcousticEquation}
\left\{
\begin{array}{l}
\rho \frac{\partial^2 u_i}{\partial t^2}  = \sum_j \sum_k e_{kij} \frac{\partial^2 \Phi}{\partial x_j \partial x_k} + \sum_j \sum_k \sum_lc_{ijkl}\frac{\partial^2 u_k}{\partial x_l\partial x_j},\\
\sum_k \left(\sum_i\sum_je_{ijk}\frac{\partial^2 u_i}{\partial x_j\partial x_k} -\sum_j\varepsilon_{kj}^S\frac{\partial^2 \Phi}{\partial x_j\partial x_k}\right) = 0, \\
c_{izkl}\frac{\partial u_k}{\partial x_l} + e_{kiz}\frac{\partial \Phi}{\partial x_k} = 0, 
\end{array}
\right.
\end{eqnarray}
\end{small}%
where the second line is the electric boundary condition with $z=0$, and the third line is the mechanical boundary condition with $z=0$, which means that the stress is free at the boundary. Then in Eq.~(\ref{con:NewtonMotion}), $T_{i3} = 0$ when $z = 0$ for $i = 1,2,3$.

Based on the model above, we have the time-dependent strain field $S_{kl}$ and the electrical potential $\Phi$ satisfying that
\begin{equation} \label{con:phiXjk}
\begin{aligned}
\frac{\partial^2 \Phi}{\partial x_j \partial x_k} = -\frac{c_{izkl}}{e_{kiz}} \frac{\partial^2 u_k}{\partial x_l \partial x_j}.
\end{aligned}
\end{equation}

The motion of the piezoelectric material reads
\begin{small}
\begin{equation} \label{con:TSEMotionEquation3}
\begin{aligned}
\rho \frac{\partial^2 u_i}{\partial t^2} & = -\sum_j \sum_k e_{kij}\frac{c_{izkl}}{e_{kiz}} \frac{\partial^2 u_k}{\partial x_l \partial x_j} + \sum_j \sum_k \sum_lc_{ijkl}\frac{\partial^2 u_k}{\partial x_l\partial x_j}. \\
\end{aligned}
\end{equation}
\end{small}

Besides, Eq.~(\ref{con:TSE}) can also be written as
\begin{small}
\begin{equation} \label{con:STE}
\begin{aligned}
S_{ij} = \sum_k\sum_l s_{ijkl}T_{kl} + \sum_k d_{kij}E_k,~i,j,k,l =1,2,3,
\end{aligned}
\end{equation}
\end{small}%
where $d_{kij}$ is given according to $e_{14}$ in Eq.~(\ref{con:TSEMotionEquation3}) and can be represented with the matrix $\mathbf{d}$ with $k=1,2,3$ for different lines of the matrix as
$$\mathbf{d}=\begin{bmatrix}
0 & 0 & 0 & d_{14} & 0 & 0\\
0 & 0 & 0 & 0 & d_{14} & 0\\
0 & 0 & 0 & 0 & 0 & d_{14}\\
\end{bmatrix}.$$

In the piezoelectric material, there are only three independent elastic constants for the $c_{ijkl}$ in Eq.~(\ref{eq:AcousticEquation}) denoted as $c_{11}$, $c_{12}$, $c_{44}$~\cite{Simon1996Coupling}. Then the equation in Eq.~(\ref{eq:AcousticEquation}) can be simplified and rewritten as
\begin{small}
\begin{equation} \label{con:SimpleSAW}
\begin{aligned}
\rho \frac{\partial^2 u_x}{\partial t^2}  =  &c_{11} \frac{\partial^2 u_x}{\partial x^2} + c_{44}\left(\frac{\partial^2 u_x}{\partial y^2} + \frac{\partial^2 u_x}{\partial z^2}\right) \\
&+\left(c_{12}+ c_{44}\right)\left(\frac{\partial^2 u_y}{\partial x \partial y} + \frac{\partial^2 u_z}{\partial x \partial z}\right) + 2e_{14} \frac{\partial^2 \phi}{\partial y \partial z},
\end{aligned}
\end{equation}
\end{small}

\begin{small}
\begin{equation} \label{con:SimpleSAWuy}
\begin{aligned}
\rho \frac{\partial^2 u_y}{\partial t^2}  = & c_{11} \frac{\partial^2 u_y}{\partial y^2} + c_{44}\left(\frac{\partial^2 u_y}{\partial x^2} + \frac{\partial^2 u_y}{\partial z^2}\right) \\
&+\left(c_{12}+ c_{44}\right)\left(\frac{\partial^2 u_x}{\partial y \partial x} + \frac{\partial^2 u_z}{\partial y \partial z}\right) + 2e_{14} \frac{\partial^2 \phi}{\partial x \partial z},
\end{aligned}
\end{equation}
\end{small}

\begin{small}
\begin{equation} \label{con:SimpleSAWuz}
\begin{aligned}
\rho \frac{\partial^2 u_z}{\partial t^2}  = & c_{11} \frac{\partial^2 u_z}{\partial z^2} + c_{44}\left(\frac{\partial^2 u_z}{\partial x^2} + \frac{\partial^2 u_z}{\partial y^2}\right) \\
&+\left(c_{12}+ c_{44}\right)\left(\frac{\partial^2 u_x}{\partial z \partial x} + \frac{\partial^2 u_y}{\partial z \partial y}\right) + 2e_{14} \frac{\partial^2 \phi}{\partial x \partial y},
\end{aligned}
\end{equation}
\end{small}

\begin{small}
\begin{equation} \label{con:potential}
\begin{aligned}
\varepsilon \Delta \phi = 2e_{14} \left(\frac{\partial^2 u_x}{\partial y \partial z}+\frac{\partial^2 u_y}{\partial x \partial z} + \frac{\partial^2 u_z}{\partial x \partial y}\right),
\end{aligned}
\end{equation}
\end{small}%
where $u_x$, $u_y$ and $u_z$ represent the displacements in the three-dimensional coordinate system with the mechanical boundary condition at $z =0$ reads
\begin{small}
\begin{equation} \label{con:boundaryM}
\begin{aligned}
T_{13} = c_{44}\left(\frac{\partial u_z}{\partial x} + \frac{\partial u_x}{\partial z}\right) + e_{14} \frac{\partial \phi}{\partial y} = 0,\\
T_{23} = c_{44}\left(\frac{\partial u_z}{\partial y} + \frac{\partial u_y}{\partial z}\right) + e_{14} \frac{\partial \phi}{\partial x} = 0,\\
T_{33} = c_{11}\frac{\partial u_z}{\partial z} + c_{12}\left(\frac{\partial u_x}{\partial x} + \frac{\partial u_y}{\partial y}\right) = 0,
\end{aligned}
\end{equation}
\end{small}%
and the electrical boundary condition at $z =0$ reads
\begin{small}
\begin{equation} \label{con:boundaryE}
\begin{aligned}
e_{14}\left(\frac{\partial u_x}{\partial y} + \frac{\partial u_y}{\partial x}\right) - \varepsilon \frac{\partial \phi}{\partial z} + \varepsilon_0 k\phi = 0.
\end{aligned}
\end{equation}
\end{small}

When the sample is fabricated on the polished (100) surface, the acoustic wave propagates along the [011] direction, and is of the format~\cite{Solution011,Simon1996Coupling,CiracPRX}
\begin{small}
\begin{eqnarray}  \label{eq:uxyz011}
\left\{
\begin{array}{l}
u_x = \frac{u}{\sqrt{2}} e^{-kz+ik(x+y)/\sqrt{2}-ivkt},\\
u_y = \frac{u}{\sqrt{2}} e^{-kz+ik(x+y)/\sqrt{2}-ivkt},\\
u_z = \tilde{u}_z e^{-kz+ik(x+y)/\sqrt{2}-ivkt}.
\end{array}
\right.
\end{eqnarray}
\end{small}%

Equation~(\ref{con:potential}) reads
\begin{small}
\begin{equation} \label{con:potentialcal2}
    \varepsilon \Delta \phi=
   \begin{cases}
   2e_{14} \left(2\frac{\partial^2 u_x}{\partial x \partial z}+\frac{\partial^2 u_z}{\partial x^2}\right), &0 < z < H,\\
   0,&\mbox{else},
   \end{cases}
  \end{equation}
\end{small}%
where $H$ represents the thickness of the IDT fingers, and $\Delta \phi = \partial^2 \phi/\partial x^2 +  \partial^2 \phi/\partial y^2 + \partial^2 \phi/\partial z^2$.

Then the electrical potential can be represented as
\begin{small}
\begin{equation} \label{con:potentialSolution}
\begin{aligned}
\phi = \chi(z) e^{ik(x+y)/\sqrt{2}-ivkt},
\end{aligned}
\end{equation}
\end{small}%
where $\chi(z)$ reads
\begin{small}
\begin{equation} \label{con:zamop}
    \chi(z)=
   \begin{cases}
   \chi_p(z) + B_1e^{kz} + B_2e^{-kz}, &0 < z < H,\\
    B_3e^{kz},&z\leq0,\\
    B_4e^{-kz},&z\geq H,
   \end{cases}
  \end{equation}
\end{small}%
which is given in Ref.~\cite{Solution011} with the detailed format of $\chi_p(z)$ and $B_j$ with $j=1,2,3,4$ are to be determined parameters.

The amplitude of the electric field by SAW can be calculated as~\cite{yinPMchip}
\begin{small}
\begin{equation} \label{con:electricfield}
\begin{aligned}
\left|\vec{E}(x,y,z)\right| = \sqrt{\left(\frac{\partial \phi}{\partial x}\right)^2 + \left(\frac{\partial \phi}{\partial y}\right)^2 + \left(\frac{\partial \phi}{\partial z}\right)^2 }.
\end{aligned}
\end{equation}
\end{small}

\subsection{Electrical fields induced by the surface acoustic wave} \label{sec:threeDelelctric}
\begin{figure}[h]
\centerline{\includegraphics[width=1\columnwidth]{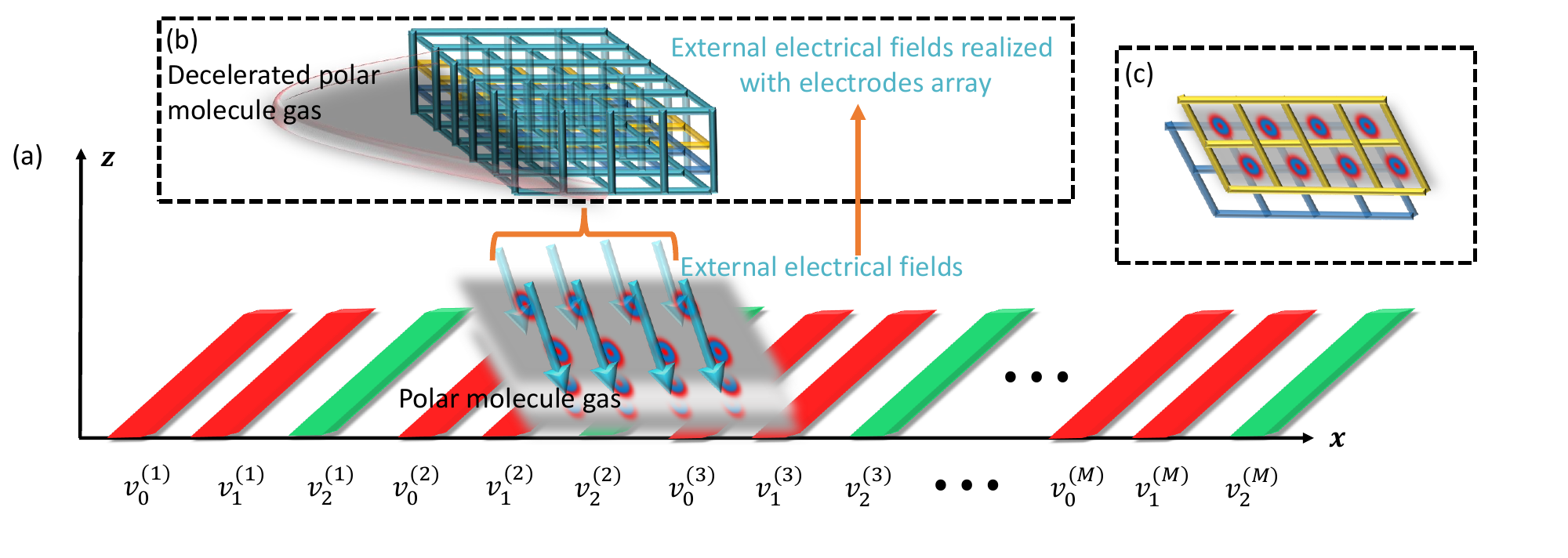}}
\caption{(a) Schematic diagram for the voltage distributions of the receiving IDT. The polar molecules interact simultaneously with both the electric field from receiving IDT and externally applied control electric field. (b) The physical realization of designed external electrical fields in (a) with an array of electrodes. (c) An example of polar molecules trapped in the lattices constructed by the electrodes.}
	\label{fig:voltage}
\end{figure}

In this section, we introduce the realization of the electrical fields interacting with the polar molecules in our proposal. As schematically shown in Fig.~\ref{fig:voltage}(a), the polar molecules can simultaneously interact with the electric field by SAW and designed external electrical fields. Experimentally, the external electric field can be realized with the three-dimensional electrode array as in Fig.~\ref{fig:voltage}(b) or Ref.~\cite{Yin2022driven}. Thus the polar molecules can be trapped in the lattices constructed and controlled by the electric fields of the electrodes such as in Fig.~\ref{fig:voltage}(c).

Based on the format of the electrical potential given in Eq.~(\ref{con:potentialSolution}) and Eq.~(\ref{con:zamop}), the polar molecule can be driven by the electrical field with the potential as
\begin{small}
\begin{equation} \label{con:FaradayGaussPotential}
\begin{aligned}
\phi =\mathfrak{R} \left[ B_4 e^{-kz} e^{ik(x+y)/\sqrt{2}-ivkt}\right],
\end{aligned}
\end{equation}
\end{small}%
which satisfies that $\Delta \phi = \partial^2 \phi/\partial^2 x^2 +  \partial^2 \phi/\partial^2 y^2 + \partial^2 \phi/\partial^2 z^2 =0$ and $\mathfrak{R}$ represents the real part of a complex value. The surface acoustic wave and polar molecules are coupled via the non-reflective IDT, as shown in Fig.~\ref{fig:TraptopDesign}. We replace $x+y$ with $x$ in Fig.~\ref{fig:voltage} and the following and the analysis in the main text for simplification, then~\cite{lin2009surface,mazalan2021current,CiracPRX}
\begin{small}
\begin{equation} \label{con:XzPotential}
\begin{aligned}
\phi(x,z,t) 
= \mathfrak{R} \left[B_4 e^{-kz} e^{ik(x-vt)}\right]= X(x,t)Z(z),
\end{aligned}
\end{equation}
\end{small}%
where
\begin{small}
\begin{equation} \label{con:PotentialXZ}
   \begin{cases}
   X(x,t)  = \mathfrak{R}\left[e^{ik(x-vt)}\right],\\
   Z(z) =  B_4 e^{-kz}.
   \end{cases}
\end{equation}
\end{small}

The period of the receiving IDT is $l = 6 p' = \lambda$ where $p'$ is the width of an electrode of the red and green fingers in Fig.~\ref{fig:voltage}. Assume the number of periods is $M$ as in Fig.~\ref{fig:voltage} and $\lambda$ is the wave length of the acoustic wave. For the $j$th finger in a period with $j =0,1,2$, when the voltage is $1$V on the electrode, the induced potential is
\begin{small}
\begin{equation} \label{con:Potentialv}
\begin{aligned}
\phi_j(x,z,t) = \sum_{m = 1}^{M} \phi_{j}^{(m)}(x,z,t),
\end{aligned}
\end{equation}
\end{small}%
where $\phi_j^{(m)}(x,z,t)$ is as the format of $\phi(x,z,t)$ in Eq.~(\ref{con:XzPotential}) and can be induced by $v_j^{(m)}$ in Fig.~\ref{fig:voltage} in proportion.

When the induced potential by the first finger of a unit in all the periods reads
\begin{small}
\begin{equation} \label{con:Potential0}
\begin{aligned}
\phi_0(x,z,t) = \sum_{m = 1}^{M} \phi_{0}^{(m)}(x,z,t),
\end{aligned}
\end{equation}
\end{small}%
then the potential induced by the other two fingers in a period reads
\begin{small}
\begin{equation} \label{con:PotentialXZ2}
   \begin{cases}
   \phi_1(x,z,t)  &= \phi_0\left(x-2p',z,t\right)\\
   &= \sum_{m = 1}^{M} \phi_{0}^{(m)}\left(x-\frac{\lambda}{3},z,t\right),\\
   \phi_2(x,z,t)  &= \phi_0\left(x-4p',z,t\right) \\
   &= \sum_{m = 1}^{M} \phi_{0}^{(m)}\left(x-\frac{2\lambda}{3},z,t\right).
   \end{cases}
  \end{equation}
\end{small}%
Assume the applied voltage on the $j$th electrode in a period is $V_j = v_j^{(1)}= v_j^{(2)} = \cdots = v_j^{(M)}$ in Fig.~\ref{fig:voltage} with $j =0,1,2$, then the overall acquired potential is
\begin{small}
\begin{equation} \label{con:TotalPotential}
\begin{aligned}
V(x,z,t) &=  \sum_{j=0}^2 V_j v_j\left(x,z,t\right)\\
&=V_0  \sum_{m = 1}^{M} \phi_{0}^{(m)}\left(x,z,t\right) + V_1 \sum_{m = 1}^{M} \phi_{0}^{(m)}\left(x-\frac{\lambda}{3},z,t\right) \\
&~~~~+ V_2 \sum_{m = 1}^{M} \phi_{0}^{(m)}\left(x-\frac{2\lambda}{3},z,t\right ),
\end{aligned}
\end{equation}
\end{small}%
where $\phi_{0}^{(m)}(x,z,t) = \mathfrak{R} \left[B_0 e^{-kz} e^{ik(x-vt)}\right]$ with $B_0$ representing the amplitude of the potential at $z=0$.

Then the induced electric field vector $\vec{E}(x,z,t)$ by $V(x,z,t)$ is
\begin{small}
\begin{equation} \label{con:ElectricalFieldVector}
\begin{aligned}
\vec{E}(x,z,t) &= - \nabla V(x,z,t) = \left[E_{x},E_z\right ]^T,
\end{aligned}
\end{equation}
\end{small}%
where
\begin{small}
\begin{equation} \label{con:Exyvector}
   \begin{cases}
   E_{x} &=- V_0 \frac{\partial \sum_{m = 1}^{M} v_{0}^{(m)}(x,z)}{\partial x} - V_1 \frac{\sum_{m = 1}^{M} v_{0}^{(m)}\left(x-\frac{\lambda}{3},z\right)}{\partial x} \\
   &~~- V_2 \frac{\partial \sum_{m = 1}^{M} v_{0}^{(m)}\left(x-\frac{2\lambda}{3},z\right)}{\partial x},\\
   E_z &=- V_0 \frac{\partial \sum_{m = 1}^{M} \mathfrak{R} \left[B_0 e^{-kz} e^{ik(x-vt)}\right]}{\partial z}\\
    &~~- V_1 \frac{\partial\sum_{m = 1}^{M} \mathfrak{R} \left[B_0 e^{-kz} e^{ik\left(x-\frac{\lambda}{3}-vt\right)}\right]}{\partial z} \\
    &~~- V_2 \frac{\partial \sum_{m = 1}^{M} \mathfrak{R} \left[B_0 e^{-kz} e^{ik\left(x-\frac{2\lambda}{3}-vt\right)}\right]}{\partial z}.
   \end{cases}
  \end{equation}
\end{small}%
For the passive receiving IDT in Fig.~\ref{fig:voltage}, $V_0 = V_1 \neq V_2$, the above electrical field can be simplified as
\begin{small}
\begin{equation} \label{con:ExyvectorSimple}
   \begin{cases}
   E_{x}&=M(V_0-V_2) B_0 k e^{-kz}  \sin\left[k\left(x-\frac{2\lambda}{3}-vt\right)\right] ,\\
   E_z&=M (V_2-V_0) B_0 k e^{-kz} \cos \left[k\left(x-\frac{2\lambda}{3}-vt\right)\right].\\
   \end{cases}
  \end{equation}
\end{small}

\section{Simplify the interaction between the polar molecule and the electric field by SAW with rotating wave approximation} \label{sec:ApendixRWA}
Here we introduce in detail on the method to simplify the dependence on the time-varying components in the Hamiltonian in Eq.~(\ref{con:Hsplit}) via rotating wave approximation. We firstly define a transformed quantum state ket $\left |\psi'(t)\right \rangle$ based on the original ket $\left |\tilde{\psi}(t)\right \rangle$, then we have~\cite{hughes2020robust,tai1988ac}
\begin{small}
\begin{subequations} \label{con:stateTransformation}
\begin{numcases}{}
i\hbar \frac{d}{d t} \left |\tilde{\psi}(t)\right \rangle = H \left |\tilde{\psi}(t)\right \rangle,\label{SeqOrig}\\
\left |\psi'(t)\right \rangle = e^{i\mathbf{A}t} \left |\tilde{\psi}(t)\right \rangle ,\label{transformStatebase} \\
i\hbar \frac{d}{d t} \left |\psi'(t)\right \rangle = H_{\omega} \left |\psi'(t)\right \rangle,\label{SeqRotate}
\end{numcases}
\end{subequations}
\end{small}%
where $\hbar = 1$ for simplification and $\mathbf{U} = e^{i\mathbf{A}t}$ represents the transformation between $\left |\tilde{\psi}(t)\right \rangle$ and $\left |\psi'(t)\right \rangle$, $H$ in Eq.~(\ref{SeqOrig}) is given in Eq.~(\ref{con:Hsplit}), and in Eq.~(\ref{SeqRotate})
\begin{footnotesize}
\begin{equation} \label{con:Homega}
\begin{aligned}
&H_{\omega} = e^{i\mathbf{A}t} \left(H - \mathbf{A} \right) e^{-i\mathbf{A}t}\\
&=e^{i\mathbf{A}t} \begin{bmatrix}
    \omega_1 -kv/2  & -|\vec{\mu}| E_z^{(j)}(x,z,t)\frac{m\Omega}{J'\left(J'+1\right)} \\
     -|\vec{\mu}| E_z^{(j)}(x,z,t)\frac{m\Omega}{J'\left(J'+1\right)} & \omega_2 + kv/2\\
  \end{bmatrix}e^{-i\mathbf{A}t}\\
&=\begin{bmatrix}
    \bar{\omega} + \frac{\bar{E}_{\Lambda}-kv}{2}  & - \frac{|\vec{\mu}|m\Omega}{J'\left(J'+1\right)}E_z^{(j)}(x,z,t)e^{ikvt} \\
     - \frac{|\vec{\mu}| m\Omega}{J'\left(J'+1\right)}E_z^{(j)}(x,z,t) e^{-ikvt} & \bar{\omega} - \frac{\bar{E}_{\Lambda}-kv}{2}\\
  \end{bmatrix},  
\end{aligned}
\end{equation}
\end{footnotesize}%
where $\mathbf{A} = \begin{bmatrix}
kv/2 & 0 \\
0 & -kv/2 
\end{bmatrix}$, $\bar{\omega} = \left(\omega_1 + \omega_2 \right)/2$, and the time-varying component can be simplified as
\begin{equation} \label{con:Homega12}
\begin{aligned}
&~~~~ -|\vec{\mu}| \frac{m\Omega}{J'\left(J'+1\right)}E_z^{(j)}(x,z,t)e^{-ikvt}\\
&= (-1)^{j} \frac{|\vec{\mu}|m\Omega M \bar{u}_j k}{J'\left(J'+1\right)}  e^{-k\tilde{z}_j}\\
&~~~~ \frac{e^{i \left[k\left(x-\frac{2\lambda}{3}-vt\right)\right]}+ e^{-i \left[k\left(x-\frac{2\lambda}{3}-vt\right)\right]}}{2} e^{-ikvt}\\
&\approx (-1)^{j} \frac{|\vec{\mu}|m\Omega M \bar{u}_j k}{2J'\left(J'+1\right)}  e^{-k\tilde{z}_j}e^{-ik\left(x-\frac{2\lambda}{3}\right)},
\end{aligned}
\end{equation}
by neglecting the fast oscillating terms containing $e^{- 2i kv t}$. Then we have Eq.~(\ref{con:Hsimple}) in the main text.

\section{Illustration on Eq.~(\ref{con:condition2})} \label{sec:ProofEq}
During the trapping of polar molecules, the motion of molecules is determined by the electrical force and Newtonian mechanics. In the longitudinal direction, the molecules can be stable because the direction of the joint force by the IDT induced electrical field and the external electrical force are different at the upper side and the lower side of the trapped points and this can provide the restoring force. In the horizontal direction, once the polar molecules can escape from the trapped localization, their motion is only determined by the IDT induced electrical field as Eq.~(\ref{con:forceCO}) in the main text. Consider the property of the electrical force at a point only affected by the IDT but not by the external force, which satisfies that
\
\begin{widetext}
\begin{footnotesize}
\begin{equation} \label{con:condition2Derive}
\begin{aligned}
&~~~~\vec{\nabla} \cdot \vec{F}(x,z,t)\\
&=-\vec{\nabla} \cdot \left ( \frac{1}{2}\frac{1}{|\vec{E}|} \frac{dU}{d|\vec{E}|} \vec{\nabla} |\vec{E}|^2 \right )\\
&=- \frac{1}{2}\vec{\nabla} \cdot \left \{\frac{1}{|\vec{E}|} \frac{dU}{d|\vec{E}|} \vec{\nabla} \left [ \left (\frac{\partial \phi}{\partial x} \right )^2 + \left (\frac{\partial \phi}{\partial z} \right )^2\right ] \right \}\\
&=-\vec{\nabla} \cdot \left \{\frac{1}{|\vec{E}|} \frac{dU}{d|\vec{E}|} \left [ \left (\frac{\partial \phi}{\partial x} \frac{\partial^2 \phi}{\partial x^2} +\frac{\partial \phi}{\partial z} \frac{\partial^2 \phi}{\partial z\partial x}\right )\vec{x} 
  + \left ( \frac{\partial \phi}{\partial x} \frac{\partial^2 \phi}{\partial x\partial z} + \frac{\partial \phi}{\partial z} \frac{\partial^2 \phi}{\partial z^2}  \right ) \vec{z} \right ] \right \}\\
&=-\frac{1}{|\vec{E}|} \frac{dU}{d|\vec{E}|} \left [ \left (\frac{\partial^2 \phi}{\partial x^2} \frac{\partial^2 \phi}{\partial x^2} + \frac{\partial \phi}{\partial x} \frac{\partial^3 \phi}{\partial x^3}+\frac{\partial^2 \phi}{\partial z\partial x} \frac{\partial^2 \phi}{\partial z\partial x}  + \frac{\partial \phi}{\partial z} \frac{\partial^3 \phi}{\partial z\partial x^2}\right ) + \left ( \frac{\partial^2 \phi}{\partial x\partial z} \frac{\partial^2 \phi}{\partial x\partial z} + \frac{\partial \phi}{\partial x} \frac{\partial^3 \phi}{\partial x\partial z^2} - \frac{\partial^2 \phi}{\partial z^2} \frac{\partial^2 \phi}{\partial x^2} - \frac{\partial \phi}{\partial z} \frac{\partial^3 \phi}{\partial^2 x\partial z} \right )  \right ] \\
&=-\frac{1}{|\vec{E}|} \frac{dU}{d|\vec{E}|} \left [ \left (\frac{\partial^2 \phi}{\partial x^2} \right )^2  +   \left (\frac{\partial^2 \phi}{\partial z^2} \right )^2  + \frac{\partial \phi}{\partial x} \left (\frac{\partial^3 \phi}{\partial x^3} + \frac{\partial^3 \phi}{\partial x\partial z^2} \right ) + \frac{\partial \phi}{\partial z} \left (\frac{\partial^3 \phi}{\partial z \partial x^2} - \frac{\partial^3 \phi}{\partial x^2 \partial z} \right ) + 2 \left ( \frac{\partial^2 \phi}{\partial z\partial x} \right )^2\right ] \\
&=-\frac{1}{|\vec{E}|} \frac{dU}{d|\vec{E}|} \left [ \left (\frac{\partial^2 \phi}{\partial x^2} \right )^2  +   \left (\frac{\partial^2 \phi}{\partial z^2} \right )^2 + 2 \left ( \frac{\partial^2 \phi}{\partial z\partial x} \right )^2\right ] \\
&=-\frac{\mu_{\rm eff}^2}{\sqrt{\left(\frac{E_{\Lambda}}{2}\right)^2 + \left(\mu_{\rm eff}|\vec{E}|\right)^2}}\left [ \left (\frac{\partial^2 \phi}{\partial x^2} \right )^2 + \left (\frac{\partial^2 \phi}{\partial z^2} \right )^2 + 2 \left ( \frac{\partial^2 \phi}{\partial z\partial x} \right )^2\right ] < 0,
\end{aligned}
\end{equation}
\end{footnotesize}
\end{widetext}%
where $\phi = \phi(x,z,t)$ is given in Eq.~(\ref{con:XzPotential})~\cite{wohlfart2008stark,SAWcalculation}.

\bibliography{SAWPMTrap}
\end{document}